\documentclass[reprint,superscriptaddress, amsmath,amssymb, aps,longbibliography]{revtex4-1}
\usepackage[english]{babel}
\usepackage{color}
\usepackage{amsmath}
\usepackage{physics}
\usepackage{xr-hyper}
\usepackage[colorlinks]{hyperref}
\usepackage{diagbox}
\usepackage{multirow}
\usepackage{amsfonts}
\usepackage{amssymb}
\usepackage{latexsym}
\usepackage{graphicx}
\usepackage{adjustbox}
\usepackage{float}
\usepackage{xr}
\usepackage{subcaption}
\usepackage{caption}
\usepackage{enumitem}
\usepackage{array}
\usepackage{ytableau}
\usepackage[nodisplayskipstretch]{setspace}

\newcommand{\al}{\alpha}

\newcommand{\ep}{\epsilon}

\newcommand{\ga}{\gamma}

\begin{document}
\title{
 Avoided crossings and   dynamical tunneling close to  excited-state quantum phase transitions
}

\author{D.J.~Nader}
\email{djulian@uv.mx}
\affiliation{Facultad de F\'isica, Universidad Veracruzana, Circuito Aguirre Beltr\'an s/n, Xalapa, Veracruz 91000, Mexico}

\author{C.A.~Gonz\'alez-Rodr\'iguez}
\email{carlosgonzalez01@uv.mx}
\affiliation{ Facultad de Ingenier\'ia, Universidad Veracruzana,  Coatzacoalcos, Veracruz, M\'exico. }

\author{S.~Lerma-Hern\'andez}
\email{slerma@uv.mx}
\affiliation{Facultad de F\'isica, Universidad Veracruzana, Circuito Aguirre Beltr\'an s/n, Xalapa, Veracruz 91000, Mexico}

\begin{abstract}

Using the Wherl entropy, we study the delocalization in phase-space of energy eigenstates in the vicinity of avoided crossing  in the Lipkin-Meshkov-Glick model.  
These avoided crossing,   appearing at intermediate energies in a certain parameter region of the model, originate classically  from pairs of  trajectories lying in  different phase space regions, which contrary to the low energy regime, are not connected by the discrete parity symmetry of the model. As coupling parameters are varied,  a sudden increase of the  Wherl entropy  is observed  for eigenstates close to the critical energy of the excited-state quantum phase transition (ESQPT). This allows  to detect when an avoided crossing is accompanied by  a superposition of the  pair of classical trajectories in the Husimi functions of eigenstates. This superposition  yields  
an   enhancement of  dynamical tunneling, which  is observed by considering  initial Bloch states that  
evolve partially  into the partner region of the paired classical trajectories, thus breaking   the quantum-classical correspondence in the evolution of  observables.      

\end{abstract}

\maketitle

\section{Introduction}

Semi-classical approximations in phase-space   are very useful  to gain intuitive insights about the behaviour of  quantum systems \cite{Quan,Benenti,Kumari}.
 The  correspondence between classical trajectories   and phase-space representation of quantum states, together with the knowledge of large (or even whole) energy portions of the classical  phase-space dynamics, usually  accessible in  systems with few degrees of freedom, allow to understand many aspects of the quantum model.
 For instance, critical phenomena as excited-state quantum phase transitions (ESQPTs)\cite{Cejnar21} are associated to unstable fixed points appearing in the classical dynamics\cite{Engelhardt15, WangPerez20, Pilatowsky20}, also the exponential growth of out-of-time-ordered correlators can be understood from the classical exponential sensitivity to initial conditions\cite{Rozenbaum17}.
 
Likewise, crossings and avoided-crossings of energy levels can  be studied from a semi-classical perspective. In one-degree of freedom Hamiltonians, as the Lipkin-Meshkov-Glick (LMG) model studied here, crossings and avoided crossings may appear when there exist different classical trajectories for the same energy. According to the von Neuman-Wigner theorem \cite{vonNeu}, crossings between states belonging to the same symmetry sector are rather rare and what semi-classically  would be expected to be a crossing, becomes an avoided crossing. 
The relation of this phenomenon with tunneling was established since the  seminal papers by Landau and Zener \cite{Zener32, Landau32} and continues to be a topic of current interest \cite{Wilkinson87,Kyu-Won18,Arranz19}.

In this contribution we use semi-classical phase space methods to shed some light in the avoided crossings that appear at intermediate energies of the LMG model. Among many-body quantum systems, the LMG model has been widely used to test many-body phenomena since it is one of the simplest models that can be reduced to invariant subspaces with only  one degree of freedom. The LMG model was originally proposed to mimic  the behavior of closed shell nuclei \cite{Lipkin}. However it has been shown that it is useful in other branches  of physics like quantum spin
systems \cite{Botet}, ion traps \cite{Unanyan}, Bose-Einstein condensates in double wells \cite{Links} and cavities \cite{Chen2}, and  has also been  employed to study quantum phase transitions (QPT) \cite{LopezMoreno,Vidal,Ribeiro08} and decoherence \cite{Relano}.

The LMG  model has a discrete parity symmetry, which is spontaneously broken for large enough couplings at low energy (in some cases also at high energy).  This spontaneous breaking yields a quasi degeneracy between pairs of states with different parity\cite{Puebla13}. Classically, this is manifested as pairs of trajectories moving in different phase space regions that are connected by the parity transformation.  At the  critical ESQPT energy, the parity symmetry is restored what is expressed  classically as trajectories which  are mapped into themselves by the parity transformation.
For a particular sector of the parameter space,     classically  disconnected and no parity-related phase space regions of degenerate trajectories  appear at intermediate energy, this  manifests in the quantum model as crossings between states of different parity  and avoided crossing between states of the same parity for specific values of the coupling parameters.

We use the Wherl entropy (the Shannon entropy of the Husimi function) to measure the phase space localization of energy eigenstates and determine  its relation with the avoided crossings and the superposition of classical trajectories appearing at energies close to the ESQPT.
Other Shannon entropies in avoided crossing have been discussed in  atomic systems \cite{Gonzalez-Ferez03,YLHe15} 
and also in billiards \cite{Kyu-Won18}.
Whereas,  the Wehrl entropy itself has been used  as a reliable indicator of quantum phase transitions \cite{Romera15} and ESQPTs \cite{WangPerez20} in several models including the  LMG model  and for analyzing the transition order to chaos in the kicked Harper map\cite{Arranz19}.  A first approximation to the  study of the  Wehrl entropy at the vicinity of avoided crossing in the LMG model was reported in Ref. \cite{Romera17},  where two kind of singular behaviours were observed: a   spike-like maximum and  a sudden interchange of localization values between the states involved in the avoided crossings. In the same reference,  the relation of this behaviour with  exceptional points was also discussed.

In this paper, we also  consider  the evolution of initial   Bloch coherent state located  on one of the degenerate classical orbits at intermediate energy. With this, we establish the relation between  the Wehrl entropy sudden increase,  the enhancement of dynamical tunneling and the consequent breaking of  the quantum-classical correspondence in the evolution of several observables, as the survival probability and the  expectation value of the  population operator.

The article is organized as follows, in section II we review briefly the LMG model, its classical limit and the classification of its parameter space according to the behaviour of the energy density of states  and the different trajectories appearing in the classical limit.  In section III, the Wehrl entropy is introduced as a measure of delocalization in phase space \cite{Gnutzmann2001}   and 
our results for the eigenstates' Wehrl entropies are discussed focusing on what happens for states involved in avoided crossings.   In section IV, the consequences of the sudden increase of the Wehrl entropy for  dynamical tunneling and breaking of the classical-quantum correspondence  are analyzed.  Our conclusions are given in section V.

\section{The Lipkin-Meshkov-Glick model}

The Lipkin-Meshkov-Glick Hamiltonian 
can be written in terms of  pseudo-spin operators
\begin{equation}
 \label{Ham}
 \hat{H}=\ep_0 \hat{J}_z + \frac{V}{2}\left( \hat{J}^2_++\hat{J}^2_-\right)
+\frac{W}{2}\left(\hat{J}_+\hat{J}_-+\hat{J}_-\hat{J}_+ \right)\,,
 \end{equation}
     which satisfy the SU(2) algebra, $[\hat{J}_z,\hat{J}_\pm]=\pm \hat{J}_{\pm}$ and $[\hat{J}_+,\hat{J}_-]=2 \hat{J}_z$. The 
 Hamiltonian  commutes with  the operator $\hat{J}^2$,  therefore one can easily perform exact diagonalization in the basis of eigenstates $|J m\rangle$  for a given value of $J$. The LMG Hamiltonian has a parity symmetry associated to    the operator 
 \begin{equation}
 \hat{\Pi}=e^{-i\pi(\hat{J}_z+ J\hat{1}) }
 \label{eq:OpPar}
 \end{equation}
 with eigenvalues $\pm 1$, and which is proportional to a rotation by an angle $\pi$ around the $z$-axis. 

For convenience we use the parametrization
\begin{equation}
 \ga_x=\left(\frac{2J-1}{\ep_0}\right)(W+V),\,\,\,\,\,\ga_y=\left(\frac{2J-1}{\ep_0}\right)(W-V),
\end{equation}
which allows to write the LMG Hamiltonian as
\begin{equation}
\label{HLMG}
 \hat{H}=\epsilon_0\left[ \hat{J}_z + \left(\frac{\ga_x}{2J-1}\right)\hat{J}_x^2+\left(\frac{\ga_y}{2J-1}\right)\hat{J}_y^2\right]\,,
\end{equation}
where $\hat{J}_\pm=\hat{J}_x\pm i \hat{J}_y$.
For simplicity, from now on  we set $\epsilon_0$ to the unity.

\subsection{Classical hamiltonian}

A classical Hamiltonian can be derived from the previous quantum operator by considering its expectation value respect to Bloch coherent states \cite{Ribeiro06}
\begin{eqnarray}
 \label{BlochCS}
 |\al\rangle&=& \frac{1}{(1+|\al|^2)^J} e^{\alpha \hat{J}_+}|J -J\rangle\\
 &=&\frac{1}{(1+|\al|^2)^J}\sum_{m=-J}^J\binom{2J}{J+m}^{1/2}\al^{J+m}|Jm\rangle\,, \nonumber
\end{eqnarray}
where $\al$ is a complex number that, when  defined in terms of the angular spherical coordinates  $\al=\tan(\theta/2)e^{-i\phi}$,
 leads to
 \begin{equation}
\label{Hclassic}
 H\equiv\langle \alpha|\hat{H}|\alpha\rangle=\epsilon_0 J\left[ z + \frac{\ga_x}{2}x^2+\frac{\ga_y}{2}y^2\right]\,,
\end{equation}
where  \begin{equation}z=-\cos\theta,\ \ \ x=\sin\theta\cos\phi,\ \ \ y=\sin\theta\sin\phi,
\label{eq:compon}
\end{equation}
thus defining  the surface of the Bloch sphere $z^2+x^2+y^2=1$. The number of degrees of freedom of the model is  two. In terms  of the set of canonical variables  $z$ and $\phi$, the Hamiltonian reads
\begin{equation}
\label{HclassicB}
H=\epsilon_0 J \left[z +\frac{1-z^2}{2}\left( \gamma_x \cos^2\phi +\gamma_y \sin^2\phi\right)\right].
\end{equation}
Another convenient set of  canonical variables well suited  to parametrize   the surface of the Bloch sphere is given by variables 
\begin{equation}
\label{QP}
Q=\sqrt{2(1-\cos\theta)}\cos\phi,\,\,\,P=-\sqrt{2(1-\cos\theta)}\sin\phi,
\end{equation}
which map the surface of the Bloch sphere to a disc  of radius two, $\sqrt{Q^2+P^2}\leq 2$, and 
  are related to the coherent parameter through   $\alpha=\frac{Q-i P}{\sqrt{4-(Q^2+P^2)}}$. With these coordinates, the south pole of the Bloch sphere  is mapped to the center of the disc, the north pole  to the disc's perimeter and  the sphere´s equator corresponds  to an  inner circle of radius one.

\subsection{Characterization of the LMG parameter space}

\begin{figure*}
\includegraphics[angle=0,width=\textwidth]{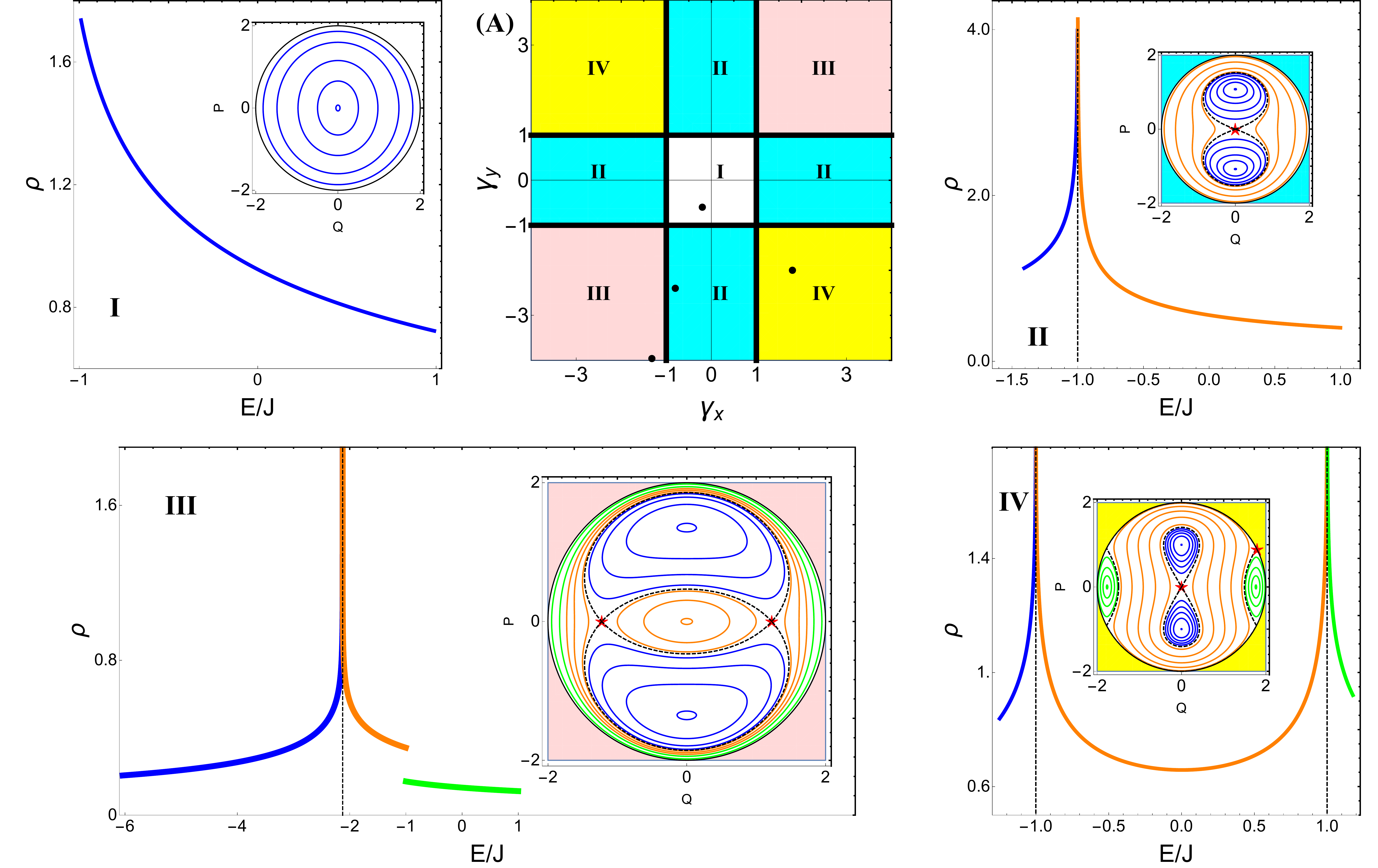}
	\caption{
	Panel (A) depicts the parameter space of the LMG model classified  according to the different  semi-classical profiles   that can be found for  the energy density of states.  Typical  energy densities of states ($\rho$) are shown in panels I-IV, the  colors of the lines indicate different energy regimes.  Inside these panels, typical classical trajectories  are  shown with colors specifying  their respective energies. In panels II-III and IV the logarithmic divergences in the energy density of states  are indicated by vertical  lines. The  separatrices in phase space associated to these divergences  are indicated by dashed lines and hyperbolic fixed points by stars. Dots in panel (A) indicate the cases shown in panels I to IV (to obtain the parameters of panel III the coordinates of the dot  have to be scaled by a factor 3).
} 
\label{figura1}
\end{figure*}

\begin{figure*}%[h]
\begin{flushleft}
\hspace{0.1\textwidth}(a)\hspace{0.35\textwidth} (b) \hspace{0.3\textwidth} (c)\\
\end{flushleft}
\includegraphics[width=.9\textwidth]{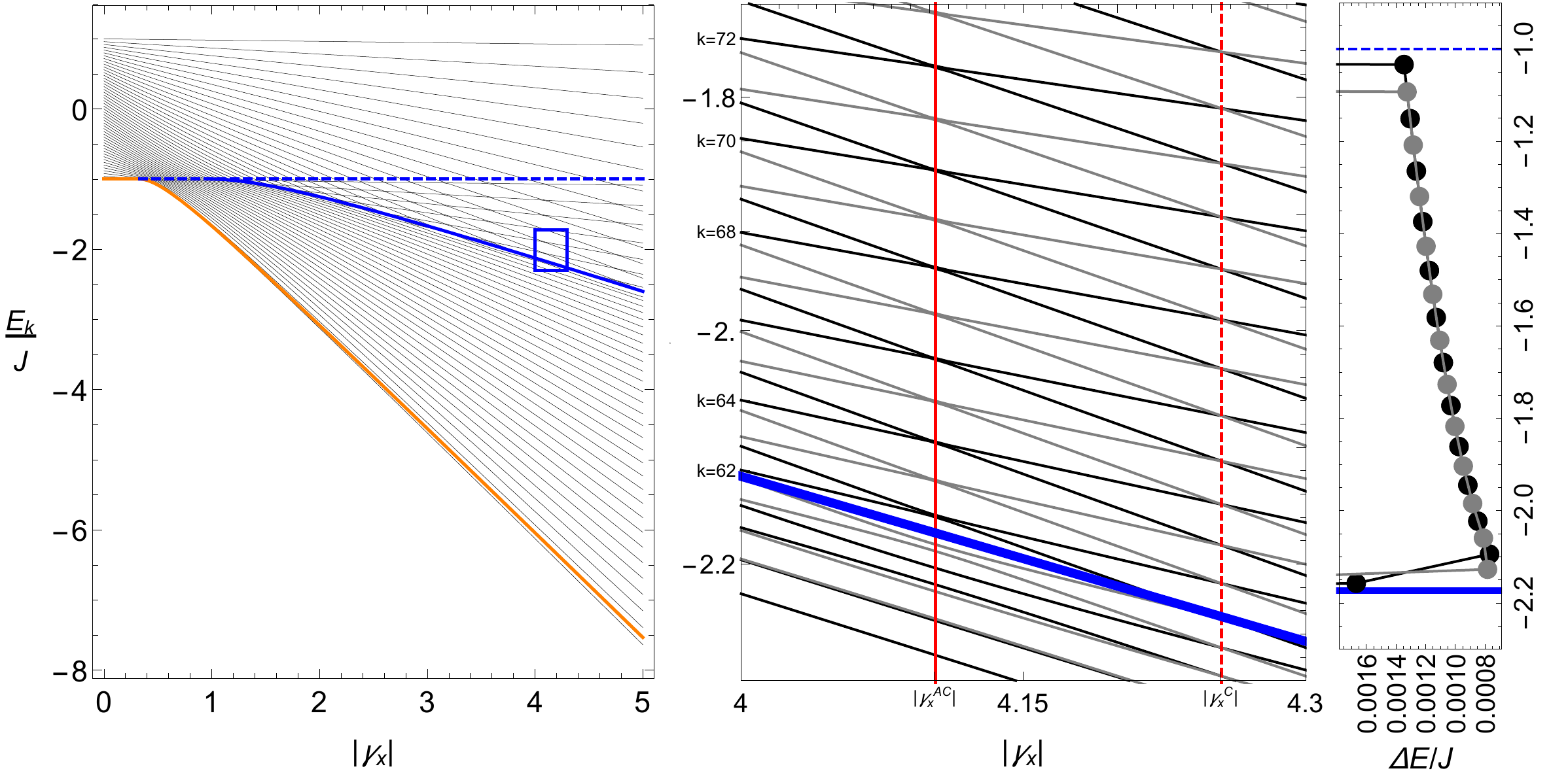}
\caption{ Energy spectrum as a function of coupling parameter $|\ga_x|$ with $\ga_x<0$ and $\ga_y=3\ga_x$.
Panel (a) shows the whole positive parity spectrum for $J=50$.  Color lines correspond to the classical energies for the ground-state (orange),  ESQPT energy (blue) and $E/J=-1$ (dashed) where a discontinuity in the EDoS occurs. Panel (b) is a zoom into the rectangle of panel (a), where avoided (vertical solid line)  and real (vertical dashed line) crossings take place.
For panel (b) a larger $J=100$ was used and the spectrum include positive parity (black lines) and negative parity (gray lines) states. Vertical lines were evaluated with Eq.(\ref{cond}) using $N=172$ (even) for the avoided-crossings  condition ($|\ga_x^{AC}|=4.10331$) and $N=173$ (odd) for the real crossings ($|\ga_x^{C}|=4.25529$). Indexes $k$ of some positive-parity states are indicated at the left axis of panel (b). Panel (c) shows the energy difference between levels in avoided crossings of positive (black dots) and negative (gray dots) parity  states along the solid vertical of  panel~(b).    
}
\label{figura2}
\end{figure*}
%%%%%%%%%%%%%%%%%%%%%%%%%%%%%%%%%%
%%%%%%%%%%%%%%%%%%%%%%%%%%%%%%%%55

The classical Hamiltonian (\ref{HclassicB}) allows to identify the phases of the  LMG model according to the properties of its ground-state. The LMG model parameter space can be classified \cite{LopezMoreno} by means of    the order parameter $\langle \hat{O}\rangle\equiv\langle \hat{J}_z+ J\hat{1}\rangle$ and the ground-state energy $E_{gs}$. For $\gamma_x>-1$ and $\gamma_y>-1$, the order parameter $\langle \hat{O}\rangle=0$ and $E_{gs}=-J$; whereas $\langle O\rangle=J(\gamma_m+\gamma_m^{-1})$ and $E_{gs}=\epsilon_0 J(\gamma_m+\gamma_m^{-1})$ for $\gamma_x\leq -1$ or $\gamma_y\leq -1$, where $\gamma_m=\min(\gamma_x,\gamma_y)$. The latter phase is characterized 
by a spontaneous breaking of the parity symmetry. By looking at  the entire spectrum of the model,  a richer phase diagram is obtained \cite{Ribeiro08}. Four different sectors  in parameter space appear,  which are characterized  by their distinctive
energy density of states (EDoS). 

The EDoS  can be approximated     semiclassically as  
\begin{equation}
\label{eq:rho}
\rho_{sc}(E)=\frac{J}{2\pi} \int dz d\phi  \   \delta \left(H(z,\phi)-E\right),
\end{equation}
 whose explicit evaluation  is presented in appendix \ref{AppA}. 
These sectors  and representative energy densities of states  are depicted in Fig.~\ref{figura1}.
The sector around the non-interacting case (  $|\gamma_x|<1$ and $ |\gamma_y|<1$, labelled as I in  Fig.\ref{figura1}) presents a simple EDoS  which behaves  monotonically as a function of  energy. The other three sectors present singularities  in the EDoS. In  sector  II ($|\gamma_x|<1$ with $|\gamma_y|<1$ or $|\gamma_y|>1$ with $|\gamma_x|<1$)   a logarithmic divergence is observed at $E/J=-1$ or  $E/J=1$ depending on the sign of the coupling with larger absolute value. In  sector IV ($|\gamma_x|>1$, $|\gamma_y|>1$ with different signs for  $\gamma_x$ and $\gamma_y$) two such divergences occur at $E/J=-1$ and $E/J=1$. Finally, at sector III ($|\gamma_x|>1$, $|\gamma_y|>1$ with same signs for  $\gamma_x$ and $\gamma_y$), besides a logarithmic divergence occurring  at $E= J(\gamma_M+\gamma_M^{-1})$ [with $\gamma_M=\text{sgn} (\gamma_x)\min(|\gamma_x|,|\gamma_y|)$],  a discontinuity is observed at $E/J=\text{sgn}(\gamma_x)$. The observed logarithmic singularities in the EDoS define what is called  Excited-State Quantum Phase Transitions (ESQPT) for one-degree of freedom systems \cite{Stransky14}.   The logarithmic divergences are consequence of hyperbolic fixed points in the underlying classical system, which in turn are associated  to separatrices of the classical dynamics. The separatrices for typical phase space portraits of sectors II, III and IV are indicated in the panels of Fig.\ref{figura1} by dashed lines. Observe that in the case of  sector III, the separatrix contains two hyperbolic fixed points, differently to the separatices of the other sectors  which possess only one.
On the other hand, the discontinuity in the EDoS observed in sector  III has to do with a local maximum (at coordinates $Q=P=0$)  in the classical energy surface which mark the end of the  family of central orange trajectories shown in panel III of Fig.\ref{figura1}.    Typical trajectories of the classical Hamiltonian are also shown for other energy regimes and regions in parameter space. The energy of each trajectory is indicated by the same colour code used for plotting the EDoS and the variables $Q$-$P$ are used to represent the Bloch sphere.

In this article we focus on the parameter sector III of  Fig.~\ref{figura1} with  $\gamma_x,\gamma_y<0$,  i.e., 
in the region where the EDoS  exhibits a logarithmic divergence at $E/J=(\gamma_M+\gamma_M^{-1})$ (with $\gamma_M=-\min(|\gamma_x|,|\gamma_y|)$ 
and a discontinuity at $E/J=-1$. 
Classical trajectories  with energies  between these two critical energies are shown by orange lines  in panel III of Fig~ \ref{figura1}.
These trajectories come in degenerate (same energy) pairs formed by  
 an inner trajectory in the center of the phase space  and  a second trajectory in the outer region.  Notice that differently to the pairs of degenerated blue trajectories at  low energy, at intermediate energies the trajectories are invariant under the parity transformation of  Eq.(\ref{eq:OpPar}) (a $\pi$ rotation around a perpendicular axis through the origin in the plane $Q$-$P$). We will show, next, that these pairs of degenerate classical trajectories are  manifested in the quantum model as avoided  and reals crossings of energy levels.

\begin{figure}
\begin{tabular}{ccc} 
% \hline
 $k=10$ &  $k=90$ & \\ 
\includegraphics[width=3.5cm, height=3.5cm,angle=0]{figures/fig3a.eps} &
 \includegraphics[width=3.5cm, height=3.5cm,angle=0]{figures/fig3b.eps} &
 \includegraphics[width=0.8cm, height=3.5cm,angle=0]{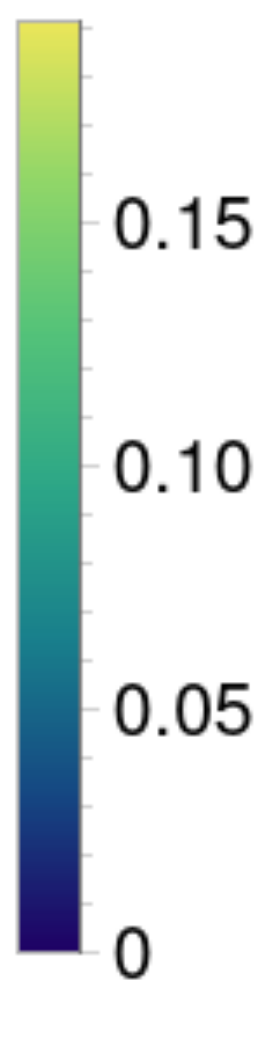}\\
 $k=64$ &  $k=65$  \\
 \includegraphics[width=3.5cm, height=3.5cm,angle=0]{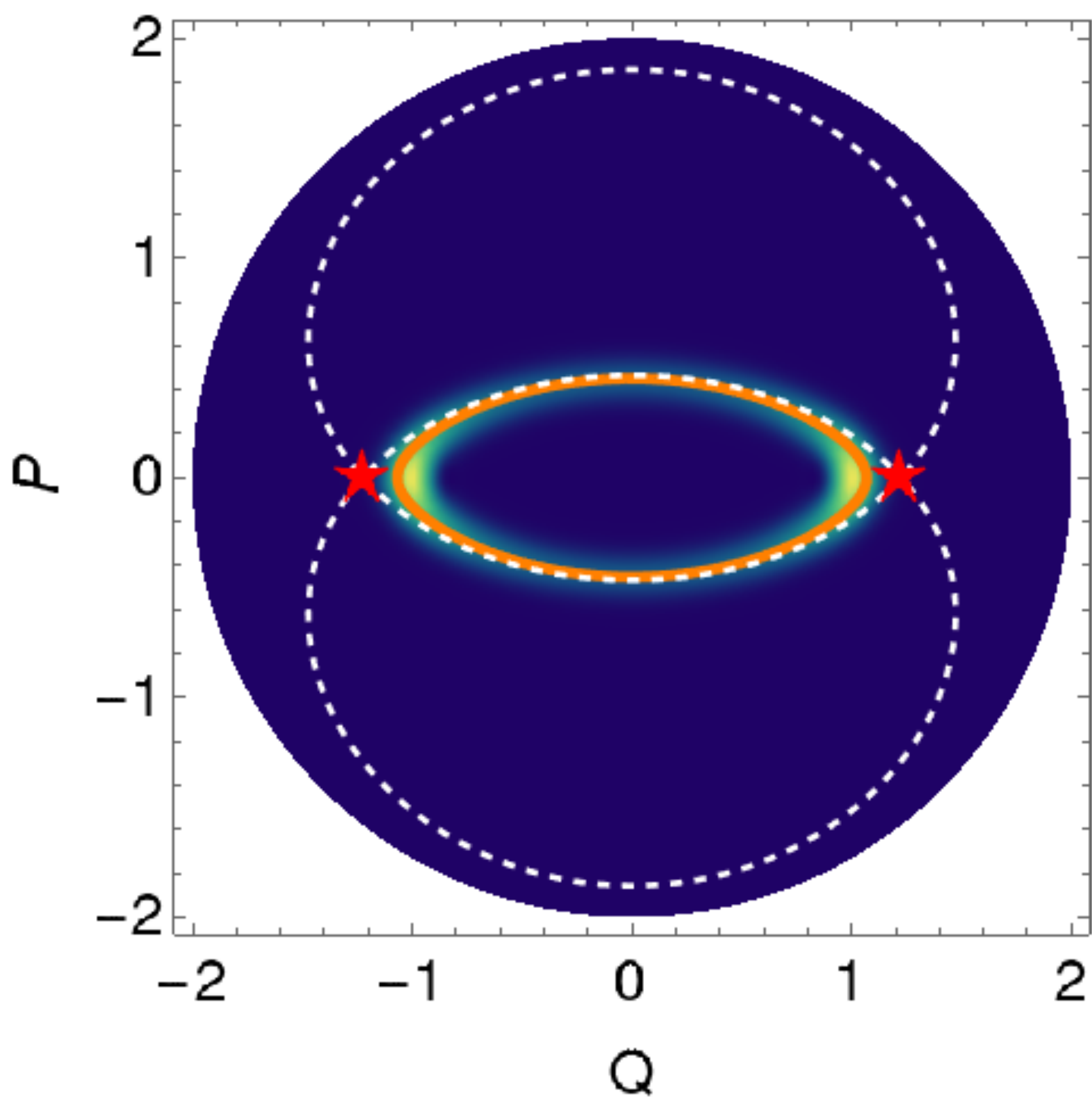} &
 \includegraphics[width=3.5cm, height=3.5cm,angle=0]{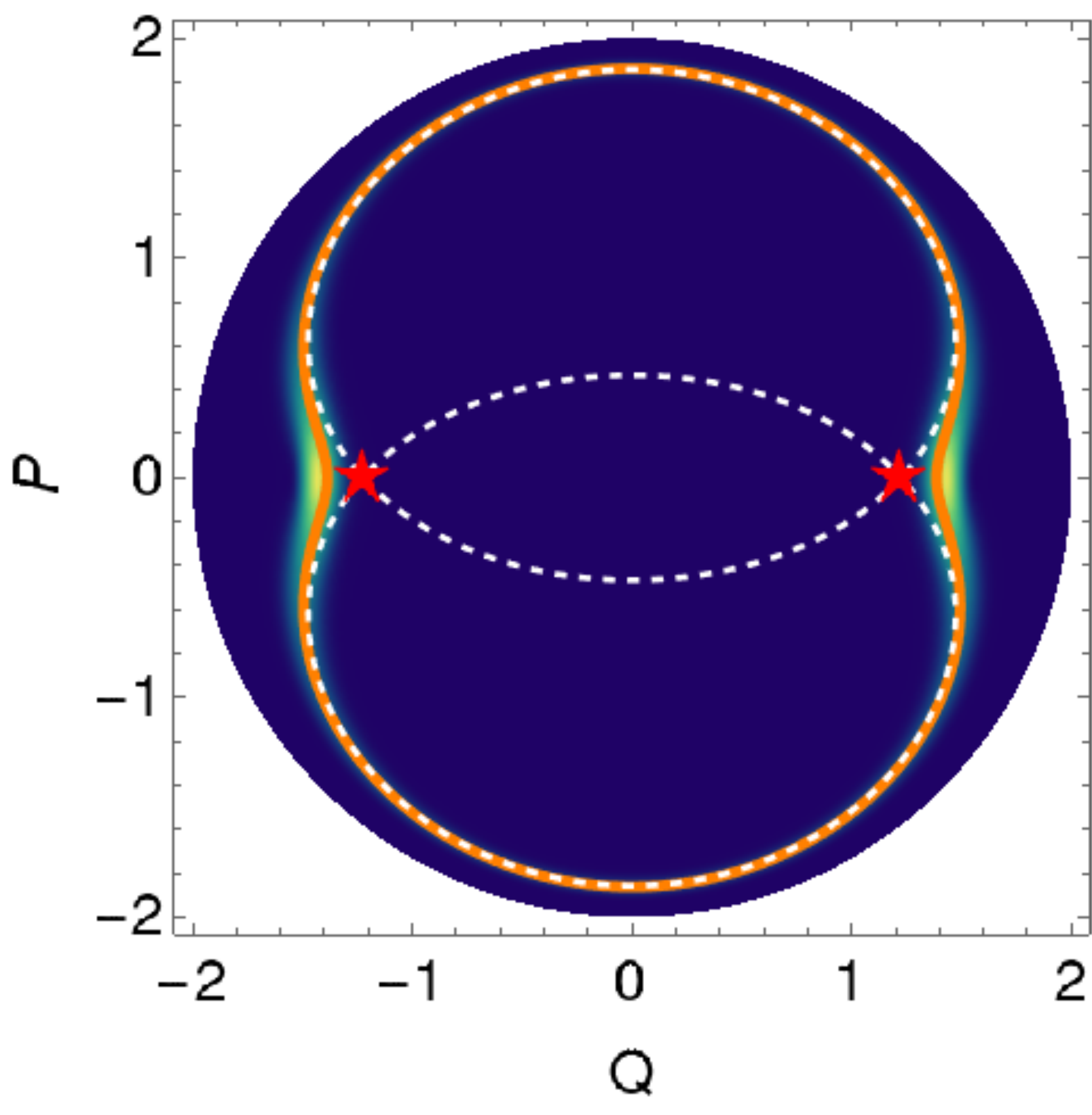} &
 \includegraphics[width=0.8cm, height=3.5cm,angle=0]{figures/fig3scale.pdf}\\
 %\hline
\end{tabular}
\begin{caption}
  {\label{Husimis} Density plots of Husimi functions $\mathcal{Q}_{k}(\alpha)$ for different eigenstates at $\ga_x=-4$ ($\ga_y=3\ga_x$) with $J=100$. Solid lines are  classical trajectories. For states $k=64$ and $k=65$, dashed lines depict classical separatrices  with associated hyperbolic fixed points indicated with stars. 
  }
 \end{caption}
 \label{figuraHusimi}
\end{figure}

\begin{figure}
\includegraphics[width=8cm, height=8.0cm]{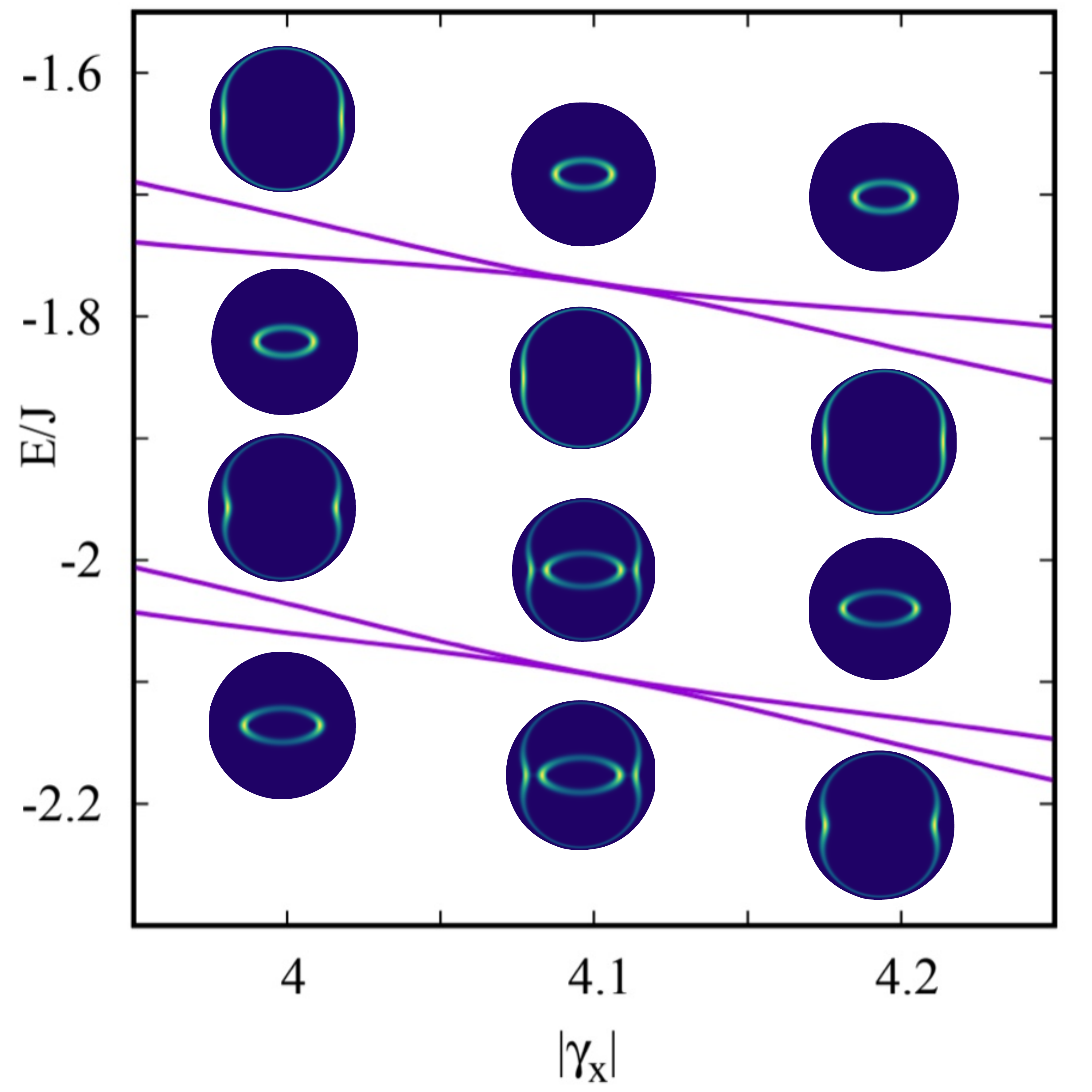}
\caption{\label{HusimiAC} Density plots  of Husimi functions $\mathcal{Q}_{k}(\alpha)$ in the vicinity of  avoided crossings ($\ga_x^{AC}=-4.10331$, with $\ga_y=3\ga_x$) between  states $k=64$-$65$ (bottom), and  $k=72$-$73$(top), located along the vertical solid line of Fig.~\ref{figura2}(b) . }
\end{figure}

\subsection{Crossings and avoided crossings}

The semi-classical formula for the EDoS (\ref{eq:rho}) describes the trend of the energy spectrum, but it is blind to details as  avoided and real crossings  that may appear in the spectrum. Fig.~\ref{figura2} shows the exact spectrum as a function of coupling $|\gamma_x|$ for $\gamma_y= 3\gamma_x$. By varying $\gamma_x$ from $0$ up to $\gamma_x=-5$, sectors I, II and  III are traversed, which is reflected by  changes in the density of states that can be appreciated in the energy levels. Panel (b) in Fig. \ref{figura2} shows a zoom into the intermediate energy interval for  couplings in the sector III of  interest. The lines clearly  show  simultaneous crossings between states with different parity (dashed vertical line) and avoided crossing  between states of the same parity (solid vertical line).  The energy gaps of the  avoided crossings,   hardly visible in panel (b),  are depicted  in panel (c), which also shows that  gaps increase as  energy does. 

From the Einstein-Brillouin-Keller (EBK) quantization rule, it is possible to determine the values of the coupling constants where crossings or avoided crossing take place at the intermediate energy regime. As mentioned above, at this energy region, there  exist, for a given energy,  two disconnected  classical trajectories $z_{\pm}(\phi,E)$. From the properties of these degenerate trajectories, we derive  (see  Appendix \ref{AppA})  the condition for  crossings (C) and  avoided crossings (AC), which  reads
\begin{equation}
 \gamma_x^{C(AC)}\gamma_y^{C(AC)}= \left(\frac{2J-1}{2J-N_{\text{o(e)}}}\right)^2,
\label{cond}
\end{equation}
with $N_{\text{o(e)}}$ an integer satisfying $0<N_{\text{o(e)}}<2J$. We have numerically verified that, indeed,  for couplings satisfying this condition,  crossings between states of different parity take place for $N_{\text {o}}$ odd, whereas for $N_{\text{e}}$ even the crossings predicted by the semi-classical EBK rule become avoided crossing between pairs of states with  the same parity. %[see Fig.\ref{figura2}(b)]. %as illustrated in Figure \ref{}.
Condition (\ref{cond}) was already reported in Ref.\cite{Castanos2005} and also in Ref.\cite{Lerma2013} from analyzing the LMG model as an integrable Richardson-Gaudin model.

In the next section  we present a phase-space study of the avoided crossings discussed here.  We will  use the Husimi function to represent the quantum states. The Husimi function allows not only  to study  the  classical-quantum correspondence but  also to measure the delocalization of the states in phase space through the  Wehrl entropy \cite{Wherl78}.

\section{Localization  measures in phase space}

%************************************************
\begin{figure*}%[h]
\begin{tabular}{cccc} 
  \includegraphics[width=4.5cm, height=4.0cm,angle=-90]{figures/fig5a.eps} &
  \includegraphics[width=4.5cm, height=4.0cm,angle=-90]{figures/fig5b.eps} &
   \includegraphics[width=4.5cm, height=4.0cm,angle=-90]{figures/fig5c.eps} &
   \includegraphics[width=4.5cm, height=4.0cm,angle=-90]{figures/fig5d.eps} 
    \\
    \includegraphics[width=4.5cm, height=4.0cm,angle=-90]{figures/fig5e.eps}& 
\includegraphics[width=4.5cm, height=4.0cm,angle=-90]{figures/fig5f.eps}&
 \includegraphics[width=4.5cm, height=4.0cm,angle=-90]{figures/fig5g.eps} &
 \includegraphics[width=4.5cm, height=4.0cm,angle=-90]{figures/fig5h.eps}
\end{tabular}
\begin{caption}
  {\label{EntropyPlots} Wherl entropy $W_{E_k}$  as a function of the coupling parameter $|\ga_x|$ ($\ga_y=3\ga_x$) for several states involved in the avoided crossings shown along the solid vertical line of Fig.~\ref{figura2}(b). The dashed vertical lines indicate the position of the avoided crossings at $|\ga_x|=4.1033108$. Different horizontal scales were used in the panels.}
 \end{caption}
 \end{figure*}
\subsection{Husimi function and localization measures}

 The Husimi representation of a pure state is nothing but its squared projection on  the coherent states (\ref{BlochCS}). 
 For eigenstates of the Hamiltonian (\ref{Ham}) they are 
\begin{equation}
\label{HusimiQ}
 \mathcal{Q}_{k}(\alpha)=|\langle \alpha | E_k \rangle  |^2\,.
\end{equation}

In Fig.~\ref{Husimis},  we show the   Husimi representation of some eigenstates using the canonical variables $Q$-$P$ of Eq.~ (\ref{QP}).  We select four positive parity eigenstates in the parameter region III of Fig.\ref{figura1} ($\gamma_x=-4$ and $\gamma_y=3\gamma_x$). The eigenstates with indexes $k=10$ and $k=90$ sit, respectively,  in the low energy region of blue trajectories and in the  high energy region of green trajectories of panel III in Fig.\ref{figura1}, while eigenstates $k=64$ and $k=65$ have intermediate energy in the region of degenerate orange trajectories. We can observe that the Husimi representations are distributed along the respective  classical trajectories corresponding to the energy of each state $H=E_k$. The intensity of the Husimi function reflects the dynamical properties of the classical trajectory: the Husimi function is more intense in the regions where the classical dynamics is slower. This is clearly seen in states $k=64$ and $k=65$ whose dynamics become slow  close to the position of  the indicated hyperbolic fixed points located, respectively,  at $(Q=\pm1.225,P=0)$.

Fig.~\ref{HusimiAC}  focuses on the Husimi functions of states involved in two different avoided crossings of those occurring along the vertical solid line in Fig.\ref{figura2} (b). 
In both avoided crossing, one can notice that there is an exchange of the Husimi representation between the two levels right after the avoided crossing. 
However, it is noticeable that in the case of the  avoided crossing with  energy closer to the ESQPT energy (bottom), 
one can observe  superpositions in  the  Husimi functions of the two avoided states. These superpositions are located atop    of the two classical trajectories with the same energy.  
Such superpositions occurs only for avoided crossings at energies close to the ESQPT critical energy. For the avoided crossings at higher energies we only observe the exchange in the Husimi representation but not the superposition at the avoided crossing, as illustrated by top avoided crossing in  Fig.\ref{HusimiAC}.

The superposition of the two classical trajectories in the Husimi functions entails a delocalization in phase space of the eigenstates involved in the avoided crossings. To measure this delocalization,  any   R\'enyi entropy in phase space  can be employed \cite{Pilatowsky21}. Here we use one of the most known, the Wehrl entropy (the R\'eny entropy of order one) which in our case is given by 
\begin{eqnarray}
 \label{EntropyW}
 W_{E_k}&=&-\int \mathcal{Q}_{k}(\al) \ln \mathcal{Q}_{k} (\al) d \Omega\nonumber\\
        & =&-\int \mathcal{Q}_{k}(Q,P)\ln \mathcal{Q}_{k}(Q,P) dQdP 
\end{eqnarray}
where $d\Omega$ is the solid  angle of the Bloch coherent states and in the second line we have written the integral in terms of variables $Q$-$P$ of Eq.~(\ref{QP}).

\subsection{Wehrl entropy and avoided crossings in the LMG model}

\begin{figure*}
\begin{tabular}{cc} 
 $\gamma_x^{AC}=-4.10331$ &  $\gamma_x=-4.15$ \\
  \includegraphics[width=.32\textwidth%, height=4.5cm
  ,angle=-90]{figures/fig6a.eps} &
 \includegraphics[width=0.32\textwidth%, height=4.5cm
 ,angle=-90]{figures/fig6b.eps} \\
\end{tabular}
\begin{caption}
  {\label{EntropyvsE} Wehrl entropy as a function of the energy $E_k$ for the whole set of  parity positive eigenstates. 
  In left panel the parameter $\gamma_x$ is chosen 
  at the avoided crossings condition  while in the right panel the parameter $\gamma_x$ is chosen away of this condition. In both cases $\ga_y=3\ga_x$ and $J=100$ as in Fig.~\ref{figura2}(b). The vertical dashed lines
  indicate the ESQPT and $E=-1$, respectively. Colors are used to distinguish states with index $k$ even (pink) and  odd (green). 
}
\end{caption}
\end{figure*}

The Wehrl entropy (\ref{EntropyW}) 
of  several pairs of states involved in avoided crossing  are plotted in Fig.~\ref{EntropyPlots}  as a function of $\gamma_x$ (with $\gamma_y=3 \gamma_x$)  for intervals around the coupling satisfying the condition for avoided crossings  $\gamma_x^{AC}\gamma_y^{AC}=\left(\frac{2J-1}{2J-N}\right)$ with $J=100$ and $N=172$.  The same coupling indicated  by  the solid vertical line in Fig.~\ref{figura2}(b).
As it was reported in \cite{Romera17} we can notice that there is an exchange in the entropy corresponding to each pair of states involved in the avoided crossings. However,  two distinctive kind of exchanges can be observed, depending on the energy of the pair of levels. For  avoided crossings with lower energy which are closer to the ESQPT critical energy, the exchange is accompanied by an abrupt rise and fall  of the Wehrl entropy.  The  local maxima   of  the  Wehrl entropy   diminish and the peaks become narrower  as we move away from the ESQPT energy and  higher-energy avoided crossings are considered. In Fig.~\ref{EntropyPlots}, the spike-like behaviour of the Wehrl entropy is clearly seen in the avoided crossing of levels $k=62$-$63$ up to $k=68$-$69$, while for the avoided-crossing of levels $k=70$-$71$ the spike is barely visible. For the rest of avoided crossing the spike has completely disappeared.    
The sudden increases in the entropy observed for the set of avoided crossings close to the ESQPT energy  are  direct manifestations of the superposition in the Husimi functions, as the one   
shown  at the  bottom  of Fig.~\ref{HusimiAC}.

In order to have a broader picture of the behaviour of the Wherl entropy, we show in Fig.~\ref{EntropyvsE} the Wherl entropies of  all the parity positive eigenstates $|E_k\rangle$  for two different couplings. Left panel corresponds to the same coupling $\gamma_x^{AC}$ considered before,  satisfying the condition of avoided crossings, and right panel is for a coupling away of this condition. To better visualize the behaviour of the Wherl entropy we use different colors for states with even and odd indexes $k$.     The behaviours of the Wherl entropy in the two couplings show some similarities: 1) for  energy states below the ESQPT energy (which is indicated by leftmost vertical dashed line), even and odd indexed states follow the same trend with Wehrl entropy increasing as a function of energy; 2) a change in the behaviour of the Wherl entropy  is observed for states approaching the  ESQPT energy; 3) for states between the two vertical lines and energy far enough of the ESQPT energy, the even and odd labelled states accommodate in two different lines  decreasing at different rates,  those with  odd indexes, whose Husimi functions concentrate around inner classical degenerate trajectories,  have lower Wehrl entropy than even indexed states, these latter ones have   Husimi functions that  concentrate around  the outer classical trajectories; 4) since the inner  trajectories disappear classically above $E/J=-1$ (energy   indicated by the rightmost vertical line), the low line of Wherl entropy values disappear for large energies, and only the upper line remains at energies above $E/J=-1$. 

However,  striking differences in the behaviour of the Wherl entropy between the two couplings can be observed for states close to the ESQPT energy.  For  coupling satisfying the condition of avoided crossings,  states with  energies  just above the ESQPT energy have higher Wehrl entropies than the corresponding states of right panel. These higher Wehrl entropy values are the same as  those in the peaks of all the plots in  top row of  Fig.~\ref{EntropyPlots} and come from the superposition of the two degenerate classical trajectories that take place in the avoided crossings.  These states with higher Wehrl entropies disappear as  energy is further increased. 
We conclude that the avoided crossings observed in the spectrum  lead to a   superposition of classical degenerate trajectories but only for a small number of states  above  the  ESQPT energy.

\section{Physical manifestations of the Wehrl entropy increase}
\subsection{Enhancement of  dynamical tunneling}

\begin{figure*}
\begin{tabular}{|cccc||cccc|}
 %&
\multicolumn{4}{c}{$\gamma_x^{AC}$}&\multicolumn{4}{c}{$\gamma_x^{C}$}\\
 \hline
 $t=0$ &  $t=2$ & $t=20$ & $t=50$ & $t=0$ &  $t=2$ & $t=20$ & $t=50$  \\ 
 \hline
(a)\includegraphics[width=1.8cm, height=1.8cm,angle=0]{figures/fig7a1.eps} &
 \includegraphics[width=1.8cm, height=1.8cm,angle=0]{figures/fig7a2.eps}  &
 \includegraphics[width=1.8cm, height=1.8cm,angle=0]{figures/fig7a3.eps} & \includegraphics[width=1.8cm, height=1.8cm,angle=0]{figures/fig7a4.eps}& (c)
 \includegraphics[width=1.8cm, height=1.8cm,angle=0]{figures/fig7c1.eps} &
 \includegraphics[width=1.8cm, height=1.8cm,angle=0]{figures/fig7c2.eps}  &
 \includegraphics[width=1.8cm, height=1.8cm,angle=0]{figures/fig7c3.eps} &\includegraphics[width=1.8cm, height=1.8cm,angle=0]{figures/fig7c4.eps}\\
 \hline
(b) \includegraphics[width=1.8cm, height=1.8cm,angle=0]{figures/fig7b1.eps} &
 \includegraphics[width=1.8cm, height=1.8cm,angle=0]{figures/fig7b2.eps}  &
 \includegraphics[width=1.8cm, height=1.8cm,angle=0]{figures/fig7b3.eps} &\includegraphics[width=1.8cm, height=1.8cm,angle=0]{figures/fig7b4.eps}& (d)\includegraphics[width=1.8cm, height=1.8cm,angle=0]{figures/fig7d1.eps} &
 \includegraphics[width=1.8cm, height=1.8cm,angle=0]{figures/fig7d2.eps}  &
 \includegraphics[width=1.8cm, height=1.8cm,angle=0]{figures/fig7d3.eps} &\includegraphics[width=1.8cm, height=1.8cm,angle=0]{figures/fig7d4.eps}\\
 \hline
 \end{tabular}
 \newline
\rotatebox{-90}{\includegraphics
[height=1\textwidth,width=.2\textwidth,angle=0]
{figures/fig7e.eps} \newline  \hspace {-2.5 cm} \rotatebox{90}{\Large{$\mathcal{L}$}}
} 
\vspace {2cm}
\caption{ \label{lineintegrals} 
Density plot of the evolution of the Husimi functions $\mathcal{Q}_{\alpha_0}(\alpha,t)$ at selected times $t$ for four different initial Bloch coherent states. Solid orange line are  classical trajectory at the same energy of the respective initial state $\langle \alpha_o| \hat{H}|\alpha_o\rangle=H$. In  panels (a) and (b) the coupling parameters are chosen at the avoided crossing condition $\gamma_x^{AC}=-4.10331$  (with $\gamma_y^{AC}=3\gamma_x^{AC}$ and $J=100$).  In panel (a) the initial state has an energy just above the ESQPT and in (b) the initial state has energy in the intermediate regime but far enough from the ESQPT. 
In  panels (c) and (d) the energies of the initial coherent states are chosen similarly to (a) and (b), respectively, but the coupling parameter is far from the condition of avoided crossings ($\gamma_x^C=-4.25529$ with $\ga_y=3\ga_x$).
Bottom row: line integral $\mathcal{L}$ of the  Husimi distribution (\ref{lineintegral}) along the partner inner classical trajectory as a function of the time for the same initial states as in panels (a)-(d).}
\label{Fig:Tunnel}
\end{figure*}

In this section we will show that the superposition of degenerate  classical trajectories that is observed in the avoided crossings of states just above the ESQPT energy yields
an enhancement of dynamical tunneling
 between regions of the phase space associated to the pair of degenerate classical trajectories.  

To this end we  consider initial Bloch coherent states centered in a point $Q,P$ on one of the degenerate classical trajectories and study its unitary evolution.  Fig.~\ref{Fig:Tunnel} shows the evolution of the Husimi function for such initial states 
\begin{equation}
\mathcal{Q}_{\alpha_o}(\alpha,t)= |\langle\alpha|\hat{U}(t)|\alpha_o\rangle|^2,
\label{evolQ}
\end{equation}
where $\hat{U}(t)$ is the time evolution operator.  Panel (a)  shows  a case when the couplings satisfy the condition of avoided crossings and the initial state has an energy  $E_{\alpha_o}=\langle\alpha_o |\hat{H}|\alpha_o\rangle$ slightly above the ESQPT critical energy. Panel (b)  shows a   case with the same couplings  but for an initial state with larger energy. Panels (c) and (d)  show similar  initials states with energies just  above the ESQPT energy and a larger one, but for  couplings not satisfying the condition for avoided crossings.  Only in the  case of panel (a), the Husimi function, initially located in the outer trajectory,  evolves into the  inner region of the phase space where the partner degenerate classical trajectory of the initial state is located. In the rest of panels, the Husimi function spreads only  along the outer trajectory of the initial state.

In summary, if the initial coherent state is located atop of one of the paired trajectories (inner/outer), its  Husimi distribution  hoop from the initial  trajectory to its pair (outer/inner) if the coupling parameters satisfy the avoided crossing condition (\ref{cond})  and if the energy of the initial state is close to  the ESQPT critical energy.
In figure \ref{Fig:Tunnel}(a), we show  the hooping from the outer to the inner region. In Appendix \ref{additional}, the complementary  case is shown,  where the hopping occurs from the inner to the outer trajectory.

In order to quantitatively confirm the dynamical tunnelling into the partner region, 
we consider the line integral of the evolved Husimi function along a classical trajectory   

\begin{equation}
 \label{lineintegral}
 \mathcal{L}(O_i,\hat{U}(t)|\alpha_o\rangle)=\oint_{O_i} \mathcal{Q}_{\alpha_o}(\alpha,t)%\rho_{|\alpha\rangle}(t)
 dl\,,
\end{equation}
where $O_i$ is the partner classical trajectory of the trajectory where the  state is initially located.
In the last row of Fig.~\ref{lineintegrals} the line integral (\ref{lineintegral}) is plotted as a function of  time for the same  initial coherent states as in panels (a) to (d).
One can notice that the line integral (\ref{lineintegral}) is significantly larger for  state (a), implying that, indeed,   dynamical tunneling occurs for  coupling parameters  at the avoided crossing condition  but only  for sates having energy close to the ESQPT. 

%%%%%%%%%%%%%%%%%%%%%%%%%%%%%%%%%%%5
\begin{figure*}
    \begin{tabular}{ccc}
     Profile & $SP$ vs time & $ j_z$ vs time\\
 (a)       \includegraphics[width=5.2cm, height=3.2cm,angle=0]{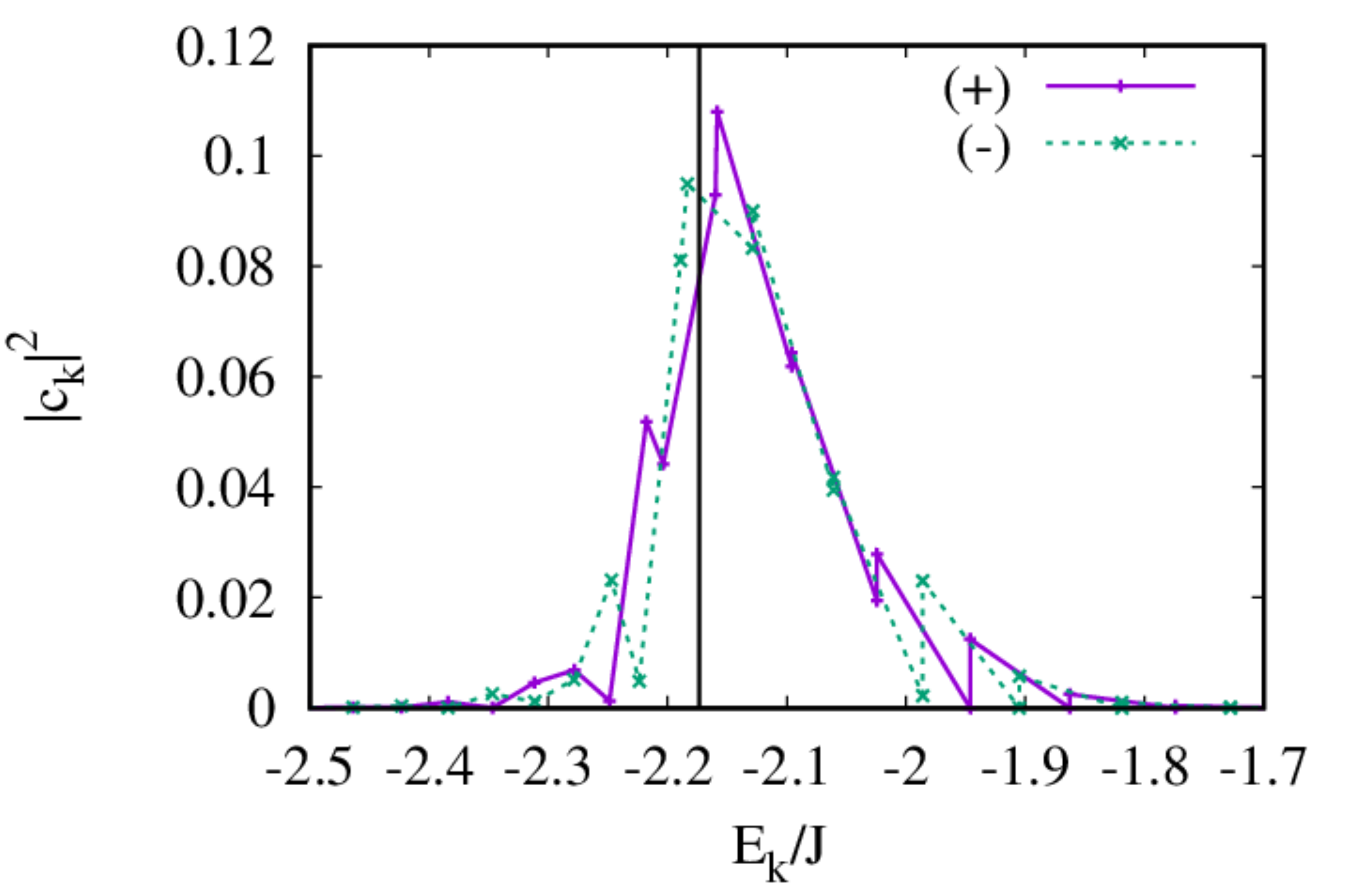} 
       &   \includegraphics[width=5.0cm, height=3.0cm, angle=0]{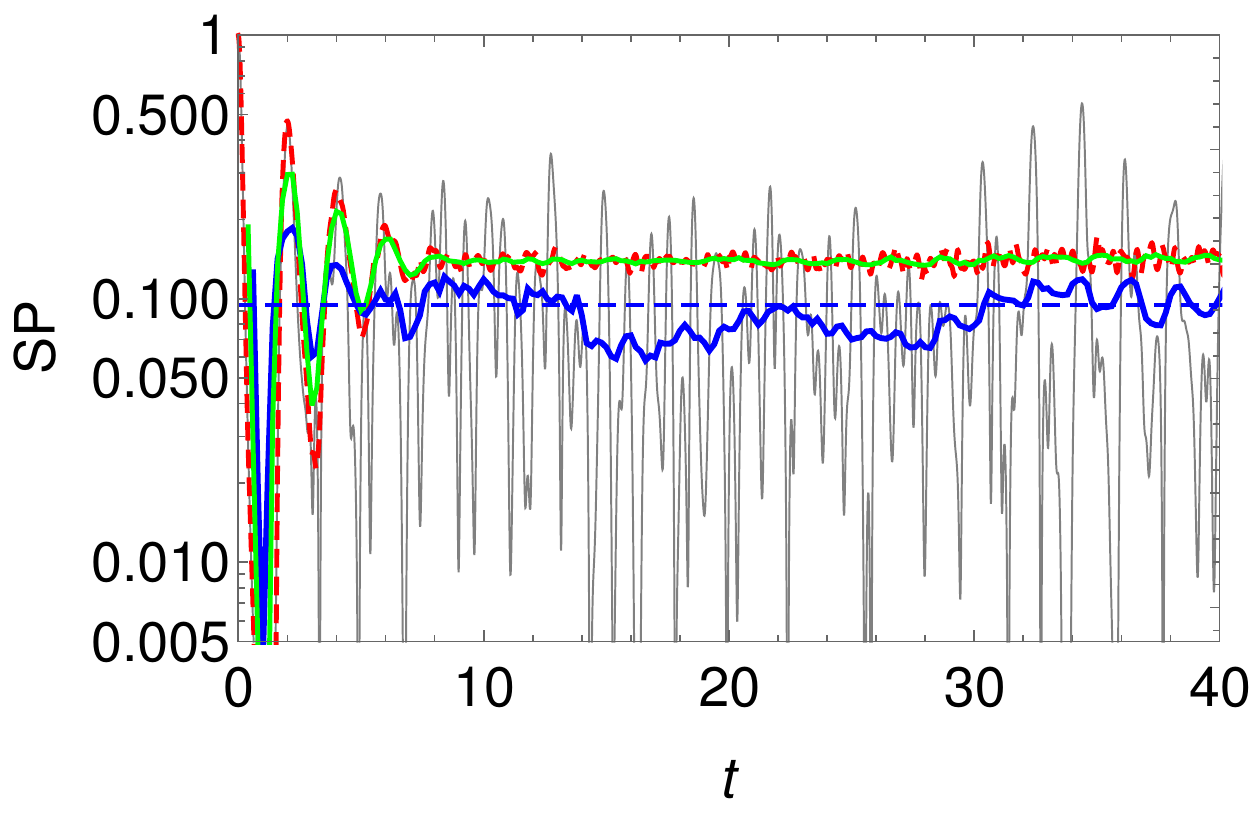}&\includegraphics[width=5.0cm, height=3.0cm, angle=0]{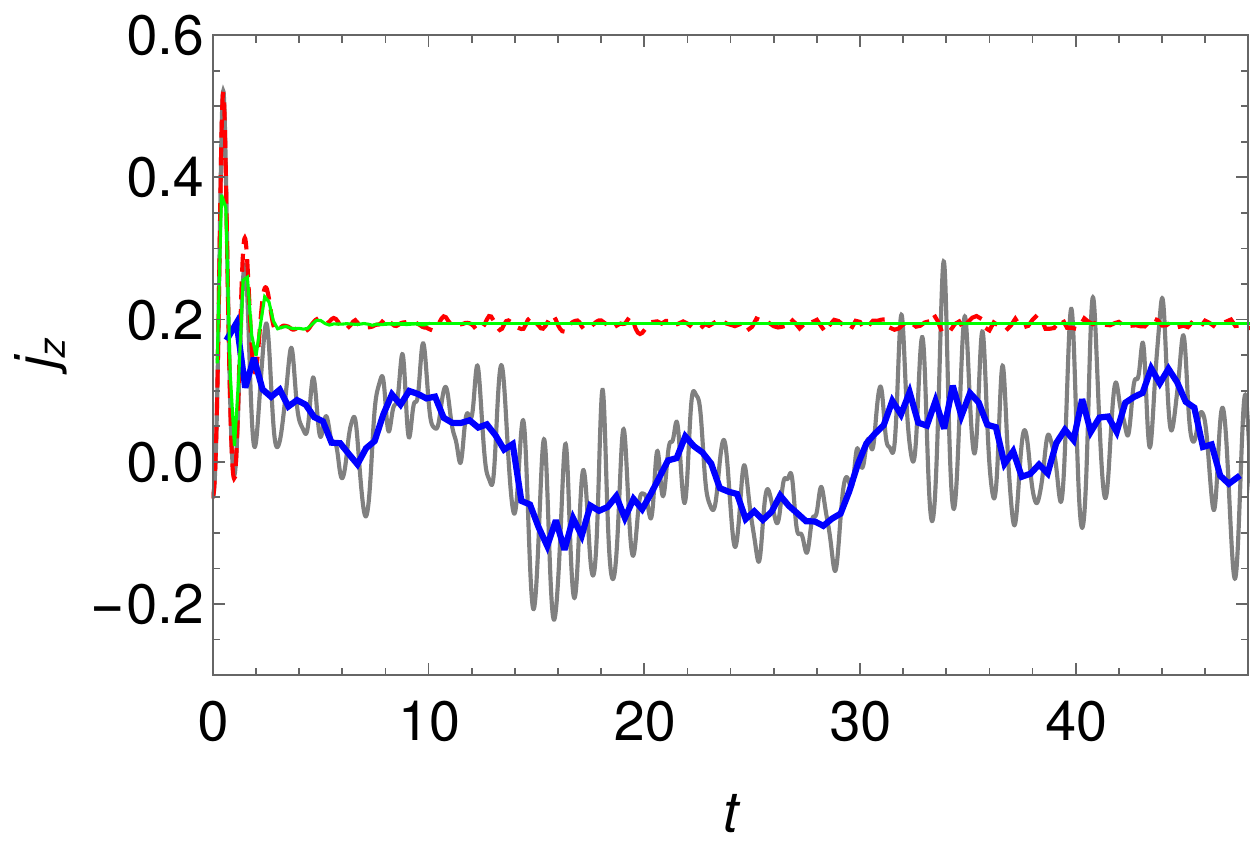}\\
       \hline
 (b)     \includegraphics[width=5.2cm, height=3.2cm,angle=0]{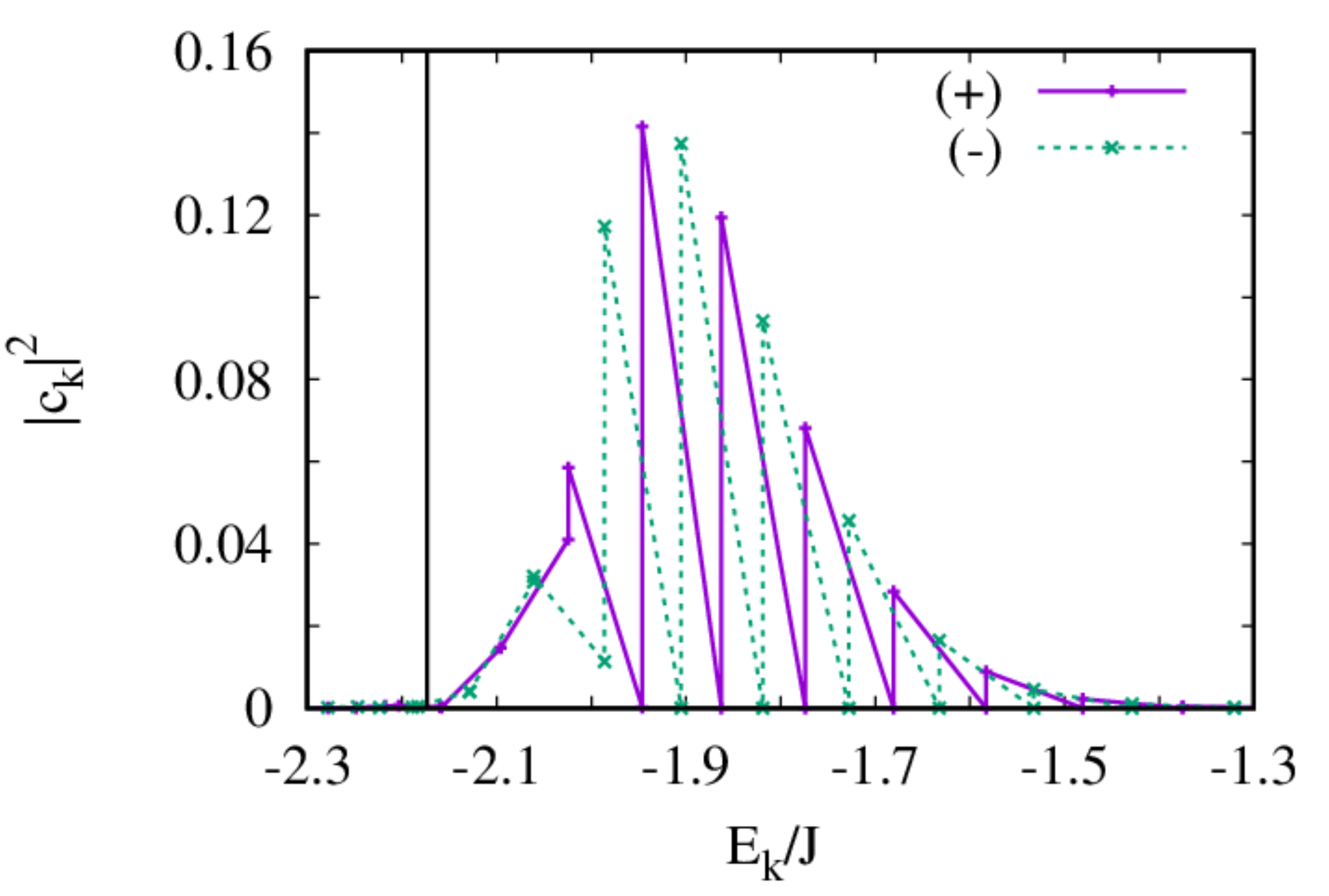} 
&\includegraphics[width=5.0cm, height=3.0cm]{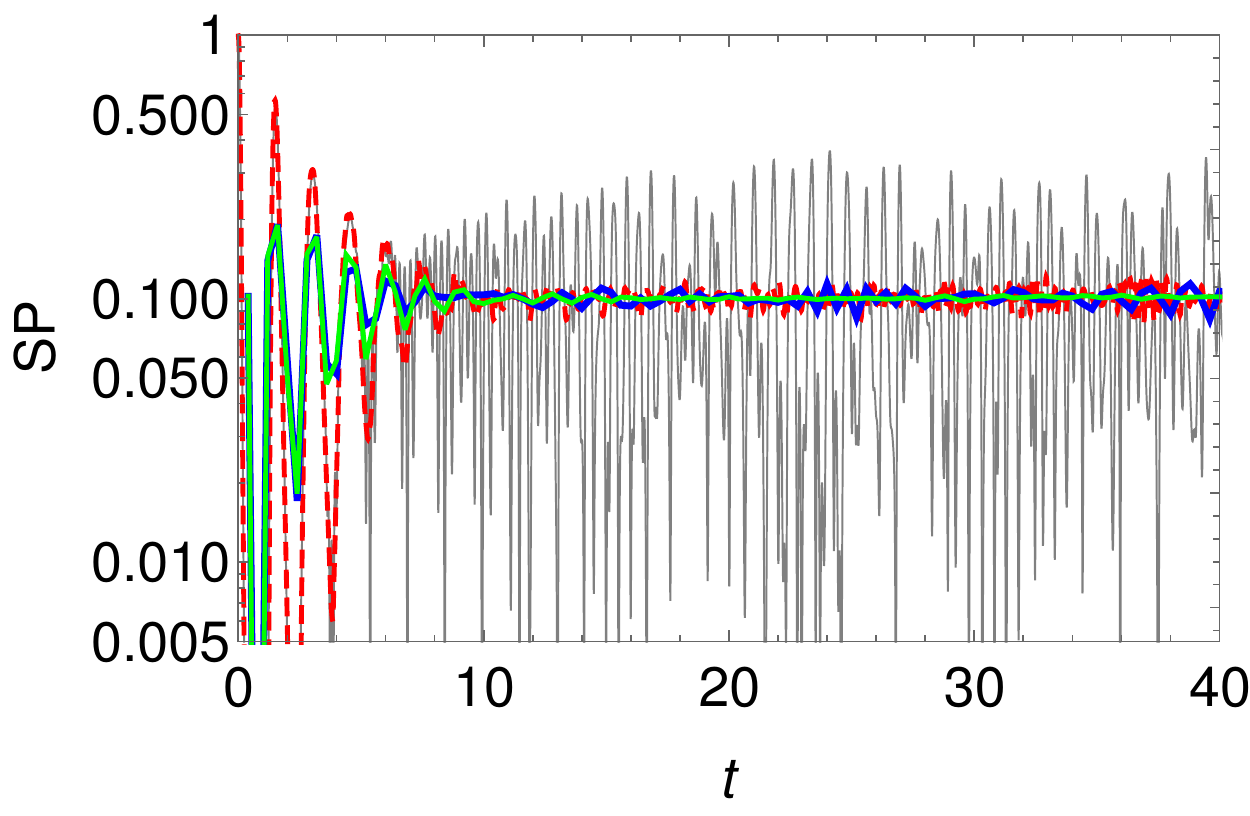}&\includegraphics[width=5.0cm, height=3.0cm, angle=0]{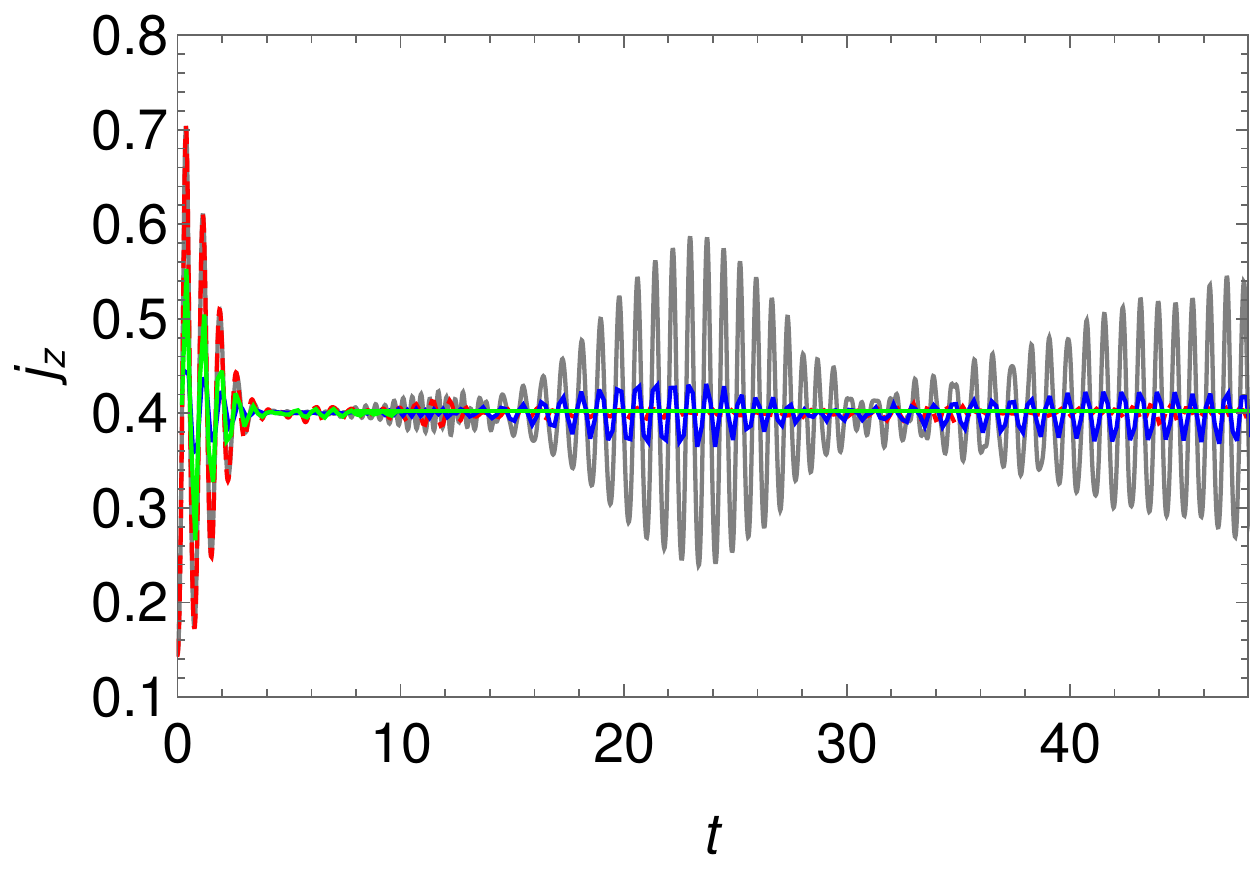} \\
\hline
\hline
(c)\includegraphics[width=5.2cm, height=3.2cm,angle=0]{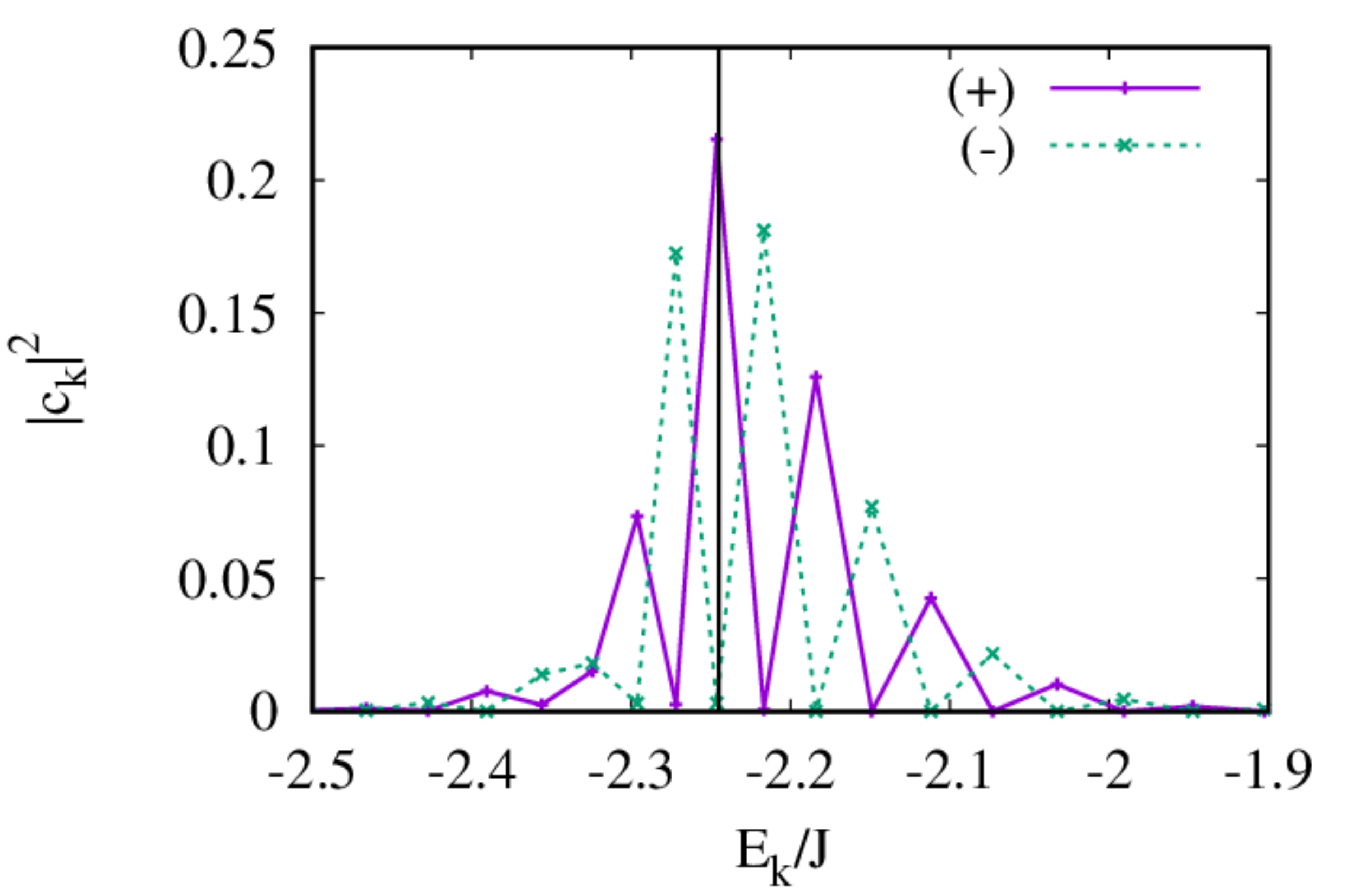} &
\includegraphics[width=5.0cm, height=3.0cm]{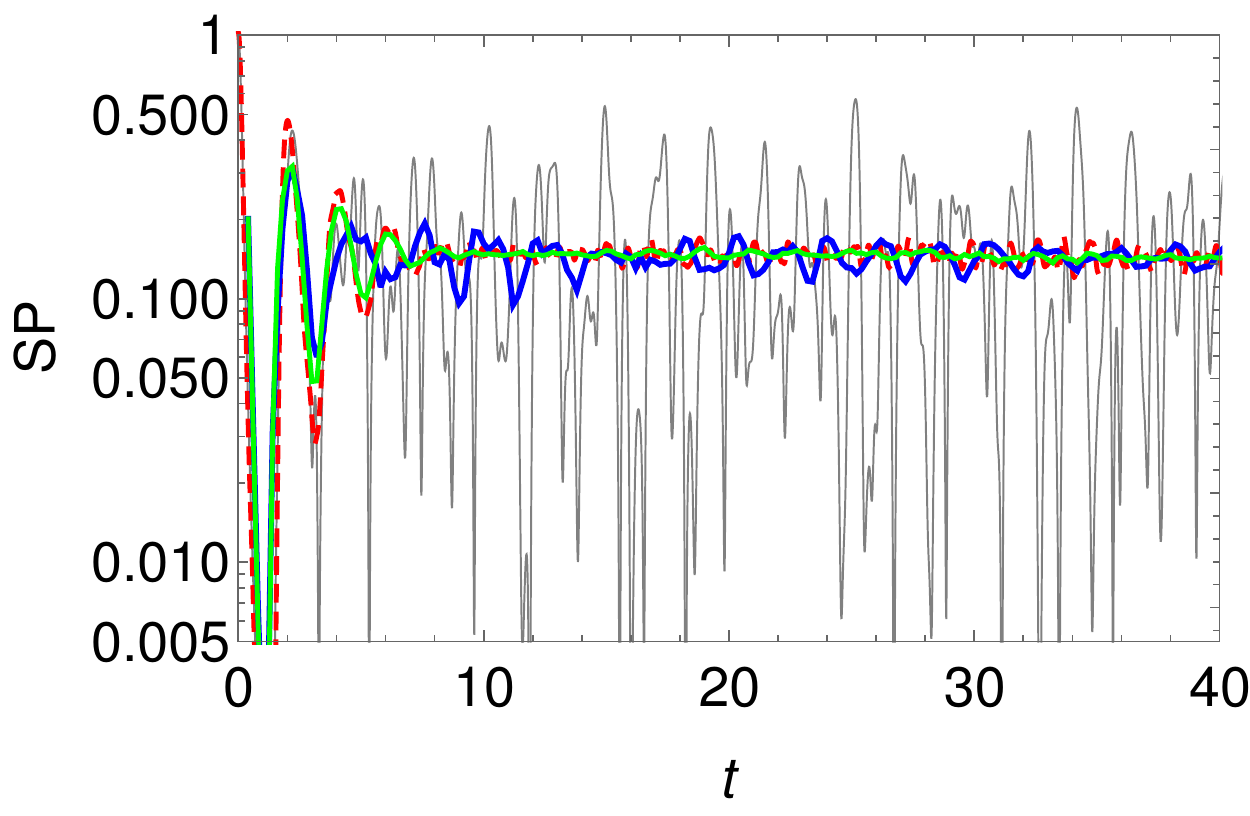}&
\includegraphics[width=5.0cm, height=3.0cm, angle=0]{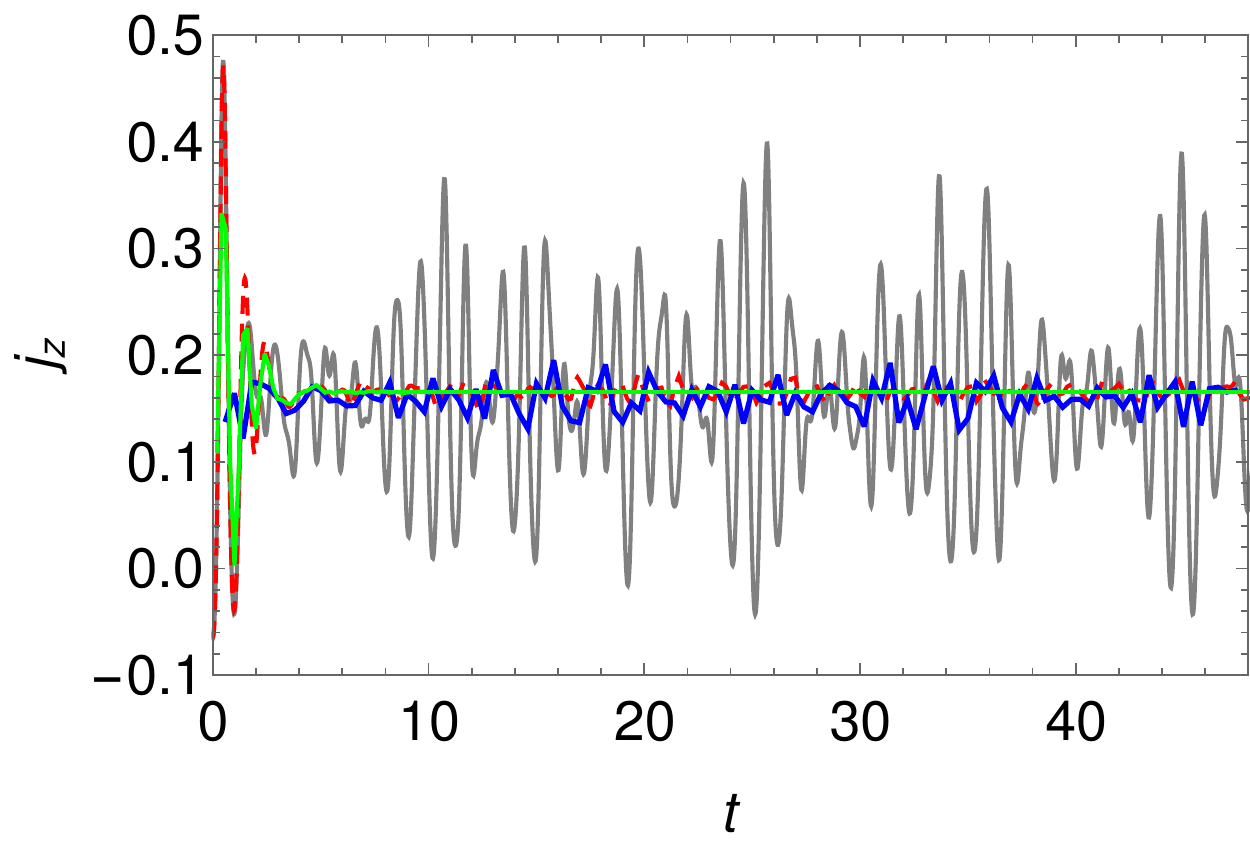} 
\\
\hline
(d)\includegraphics[width=5.2cm, height=3.2cm,angle=0]{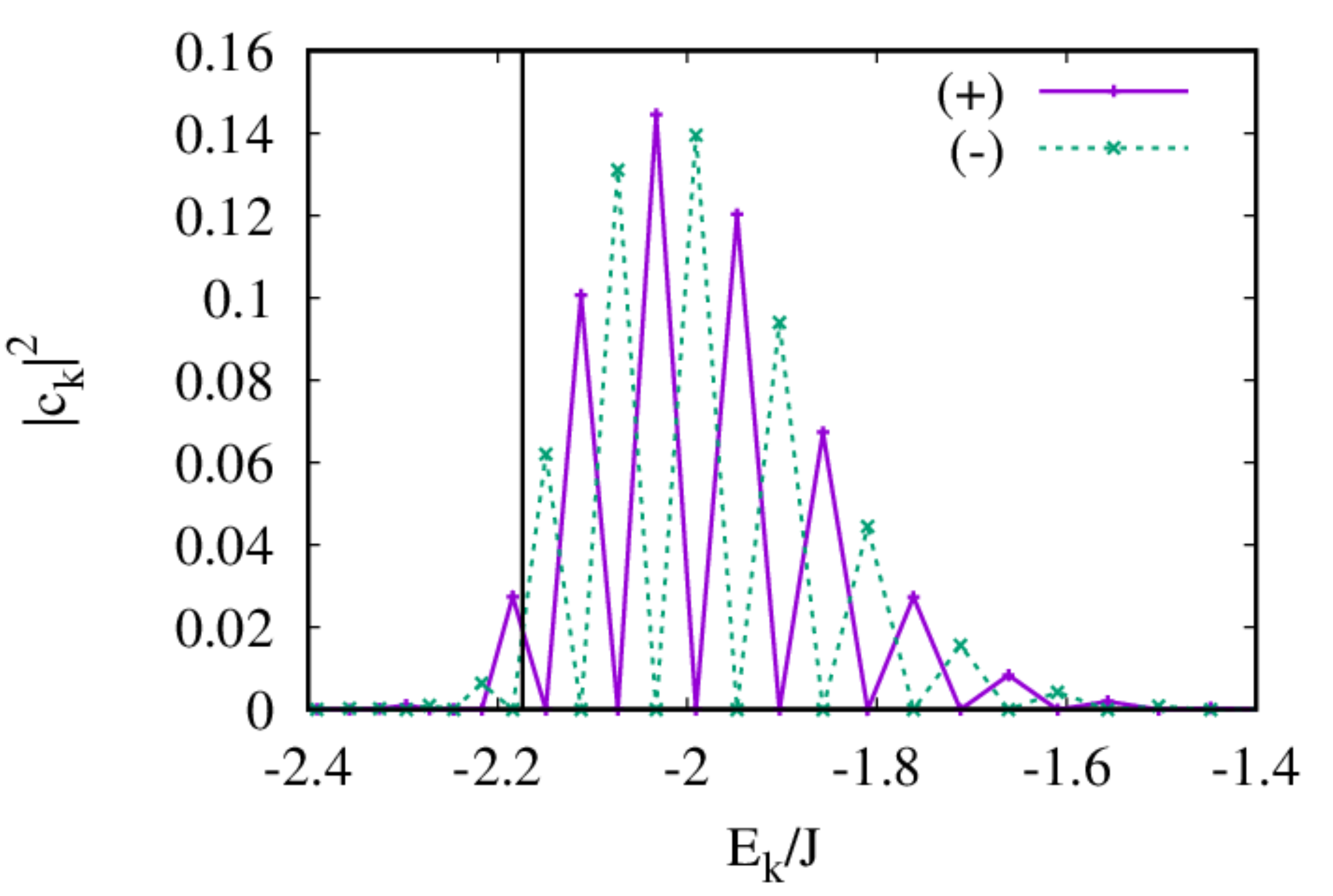}
&\includegraphics[width=5.0cm, height=3.0cm]{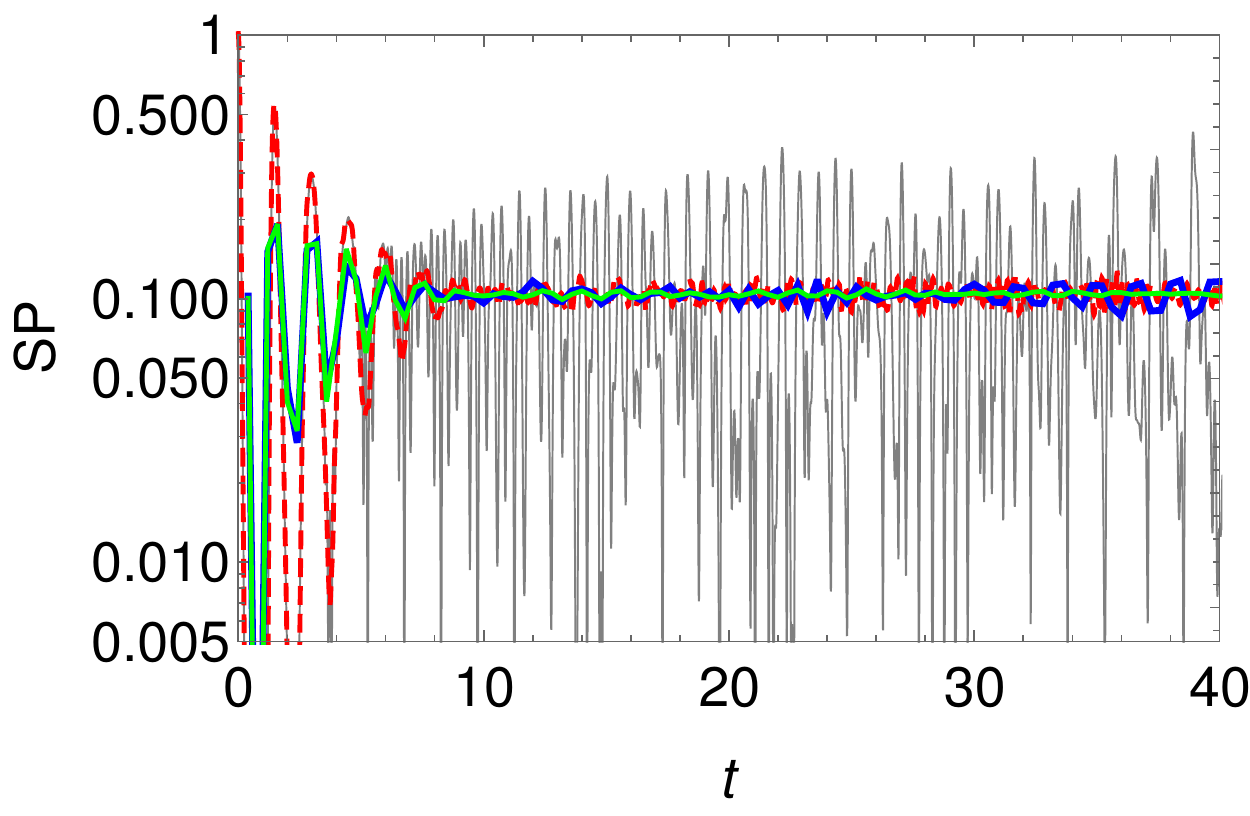}&\includegraphics[width=5.0cm, height=3.0cm, angle=0]{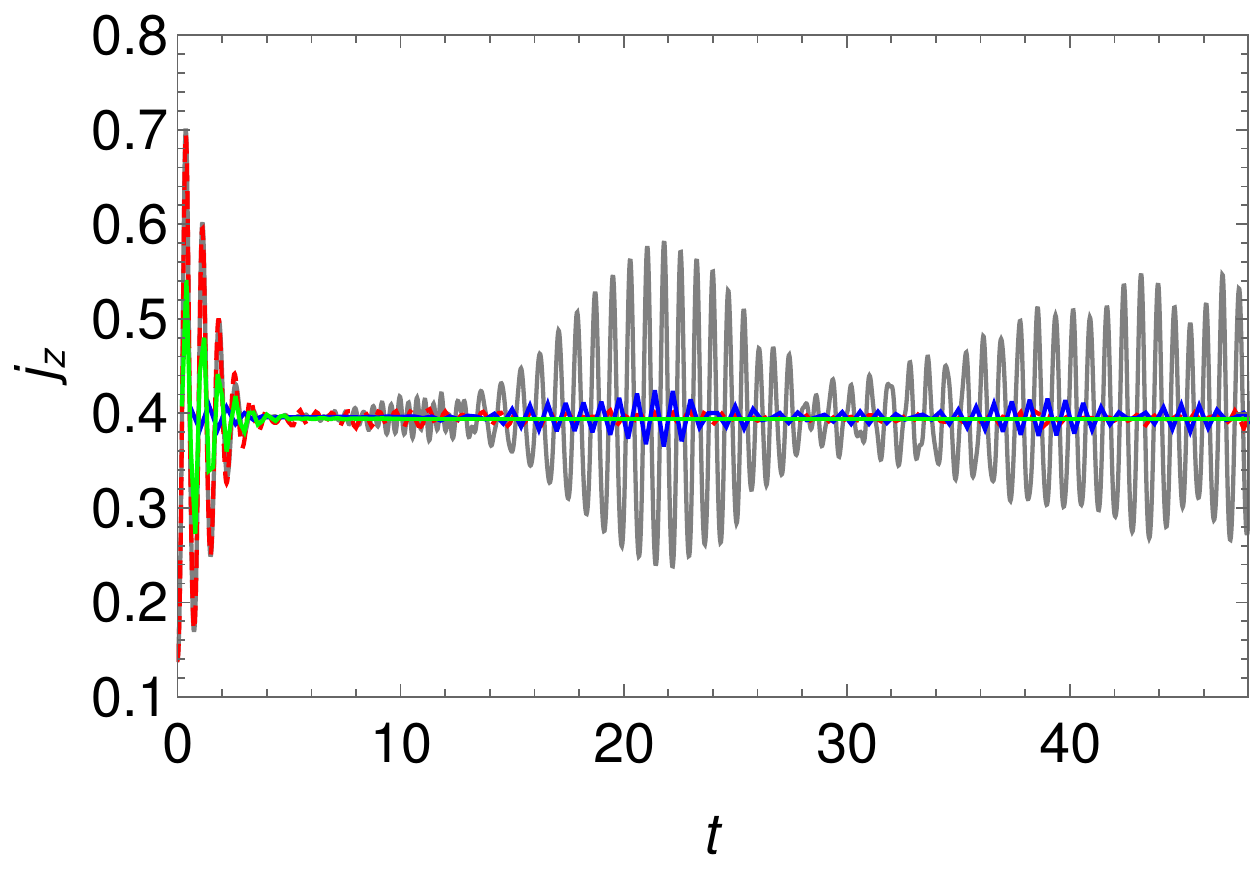} \\
\hline
    \end{tabular}
    \caption{Profiles of energy  components (left column),  survival probability ($SP$)  in logarithmic scale as a function of  time (central column) and temporal evolution of the expectation value  $j_z$ (right column) for the same   initial coherent state as in Fig.~\ref{lineintegrals}.  In left column, components from  positive parity states $(+)$ in pink solid lines and contributions from negative parity states ($-$) in dashed green lines; the position of the ESQPT is marked with a solid vertical line. 
In central and right columns, the quantum results correspond to  gray light lines and classical results to  red dashed lines. Rolling time averages are also included: blue for  quantum results and green for  classical ones.}
    \label{fig:8}
\end{figure*}
%%%%%%%%%%%%%%%%%%%%%%%%%%%%%%%%%%%%%%

\subsection{ Evolution of observables and breaking of the classical-quantum correspondence}
Being the tunneling a purely quantum effect, it may manifest as deviations in the temporal evolution of observables respect to the evolution calculated from a classical approximation. 
In this section, we will show that for initial coherent states, with energy slightly  above the ESQPT  and for  couplings satisfying the condition of avoided crossings,   the classical-quantum correspondence for  the evolution of  observables is broken.  In contrast, if the previous conditions are not fulfilled, a remarkable accord between classical and quantum  results is obtained.

As observables, we consider
 the survival probability (SP) or fidelity, 
\begin{equation}
\label{SPformula}
 SP(t)=|\langle \al_0|\hat{U}(t)|\al_0\rangle|^2,
\end{equation}
and  the temporal evolution of the population operator 
\begin{equation}
j_z(t)\equiv\frac{\langle\hat{J}_z\rangle(t)}{J}=\frac{1}{J}\langle \alpha_0|\hat{U}^\dagger(t)\hat{J}_z\hat{U}(t)|\alpha_0\rangle.
\label{eq:jzevol}
\end{equation}
As described in Ref.~\cite{Villasenor} and Appendix \ref{AppB}, both quantities  can be evaluated classically by using the classical equations %of motion and the Wigner distribution, 
through the so called Truncated Wigner Approximation (TWA). 

In Fig.~\ref{fig:8} we show the results for   couplings and same four  initial Bloch coherent states as in Fig. \ref{lineintegrals}. The first two rows correspond to  couplings satisfying the condition of avoided crossings, with the first row  showing results for  a coherent state with energy slightly above the ESQPT and  the second row for a state with larger energy.  Last two rows show results for couplings away from the condition of avoided crossings and energy just above the ESQPT critical energy and a larger one. In left column the squared energy components $|c_k|^2$  of the initial states ($|\alpha_o\rangle=\sum_k c_k |E_k\rangle$) are shown

In all  cases, quantum and classical results  for the $SP$ (central column) and $j_z$ (right column) coincide until the Ehrenfest time, after this time  the quantum results show larger oscillations than the classical approximations. However, a  fairer comparison is obtained by considering rolling temporal averages, both, for the $SP$ and $j_z$.  We observe that except in the first row, we obtain a remarkable classical-quantum correspondence, which indicates that in all these cases the dynamical tunneling is marginal and the quantum temporal trend is very well described classically.
On the contrary, for the first row 
the classical-quantum correspondence  is broken and the classical survival  probability and $j_z$ are larger than the quantum ones.
These are  additional indications that the avoided crossings favor the dynamical tunneling if  the initial coherent state is chosen  on a classical trajectory with energy slightly  above   the critical energy of the ESQPT. 

Note that in the case of the coherent state of the first row, all the  energy components  in the relevant energy interval participate in  the initial state, whereas in the other cases we obtain components very close to zero that alternate (for a given  parity) with components different to zero. This is understandable because in the case of the state of the first row,  the energy eigenstates contributing the most to the initial state are located in a energy region just above the ESQPT energy where a superposition of the  two classical trajectories is obtained. On the other hand, for the other cases, the Husimi function of the eigenstates are located either on one trajectory or other in an alternating way.

In all  cases shown in Fig.~\ref{fig:8}, the initial coherent state is located on the outer trajectory of the pair of degenerate classical trajectories, and in the case of the state of the first row the tunneling occurs into the inner trajectory of the degenerate pair. In Appendix \ref{additional}, we confirm that similar results to those in  Fig.~\ref{fig:8}  are obtained for  initial coherent states located in the inner trajectory of the degenerate pair. Likewise, in Appendix \ref{additional} we show that for initial states with energy below the ESQPT and energies above $E/J=-1$, a classical-quantum correspondence is obtained even if the couplings fulfill the condition for avoided crossings.

\section{Conclusions}

For a particular region of the parameter space in the Lipkin-Meshkov-Glick model ($\gamma_x<-1$ and $\gamma_y<-1$) there exists an intermediate energy region 
delimited by a logarithmic divergence and a discontinuity in the energy density of states, with the former defining a so-called excited-state quantum phase transition (ESQPT). At this intermediate energy region, real and avoided crossings  occur for coupling parameters fulfilling a condition that  was derived semi-classically. The real and avoided crossings were linked to the existence of pairs of different classical trajectories for the same energy that are not connected by the discrete  parity symmetry of the model. Furthermore, the ESQPT  was linked to the presence of a pair of hyperbolic fixed points in the phase space of the classical model.

We studied the avoided crossings appearing at this intermediate region  through  the Wehrl entropy, which is a measure of the phase space volume  occupied by the quantum states. We found, as a function of the coupling parameters,  a sudden increase and interchange of the Wehrl entropy for eigenstates participating in avoided crossings and having energy close to the critical ESQPT energy. For avoided crossings far enough of the ESQPT energy, the sudden increase of the Wehrl entropy disappears and only  a simple interchange is observed. 
It was shown that the spike-like  behaviour of the Wehrl entropy for avoided crossing close to the ESQPT comes  from a superposition of the degenerate classical trajectories in the Husimi (phase-space) representation of the eigenstates involved in the avoided crossings; this superposition is absent in higher energy avoided crossings. The superposition implies a sudden augmentation  of the phase space occupied by the states, and therefore an increase in the Wehrl entropy. 

The singular behavior of the Wehrl entropy in avoided crossings  for eigenstates close to the ESQPT energy enhance the dynamical tunneling between classically disconnected phase-space regions  for non-stationary states having large components of these Hamiltonian eigenstates. 
It was also shown the way as this dynamical  tunneling associated to avoided crossings induces a breaking of the quantum-classical correspondence in the temporal evolution of observables. This  quantum-classical breaking  occurs in the model, exclusively for states possessing large components of eigensates participating in the avoided crossings close to the ESQPT critical energy. For other  non-stationary states   the correspondence is kept.

The mechanism of quantum tunneling and classical-quantum breaking identified here for the Lipkin-Meshkoc-Glick model, could be also present in other quantum systems. Not only for one-degree-of-freedom models with pairs of hyperbolic fixed points, but also in models with more degrees of freedom where hyperbolic fixed points appears when the integrability of the models  is broken and they  move toward a chaotic regime \cite{Arranz98}.
The results presented here could also be  relevant in the study of dynamical phase transitions \cite{Lewis21}, where avoided crossings and closeness to hyperbolic fixed points could  influence the dynamics of non-stationary states in a similar way as we identified  here for coherent states in the Lipkin-Meshkov-Glick model.  

\section*{ACKNOWLEDGMENTS}
SL-H acknowledges useful discussions with M. Bastarrachea, J. Hirsch, S. Pilatowsky, L. Santos and D. Villase\~nor with whom several techniques used in the paper were developed. 
DJN acknowledges financial support from PRODEP project
number 42027 UV-CA-320 (Mexico),
and SL-H  from the Mexican CONACyT project CB2015-01/255702.

\appendix
%%%%%%%%%%%%%%%%%%%%%%%%%%%%%55
\begin{figure*}
I\hskip .25\textwidth  II \hskip .2\textwidth III \hskip .2\textwidth IV  \newline
 \includegraphics[width=0.95\textwidth]{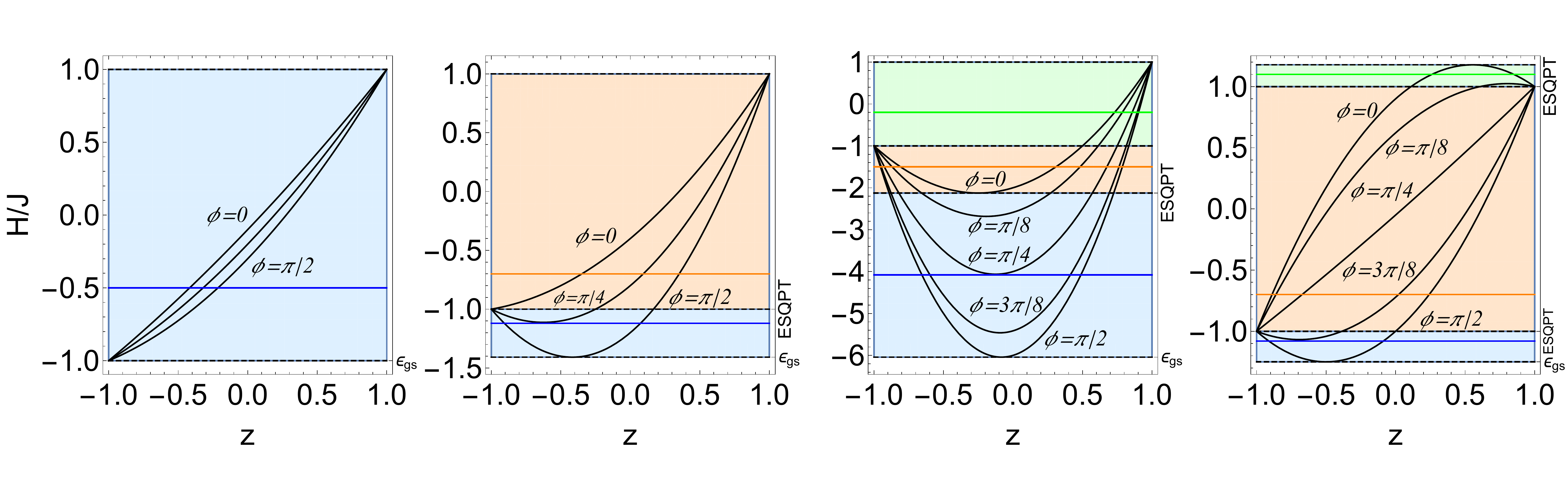}
 \caption{\label{fig:App1} Solid black lines depict the classical energy $H(z,\phi)/J$ plotted as a function of $z$ for representative values of  $\phi\in [0,\pi/2]$. The value $\phi$ used in each line is indicated, but each line is actually associated  to  four different angles $(\pm\phi, \pi\pm\phi)$.  The same sets of couplings $(\gamma_x,\gamma_y)$ as in Fig.\ref{figura1} were used, which are representative of the sectors I-IV in parameter space.  Horizontal solid lines are representative  energies of the different  regimes, classified according to the kind of trajectories $z(E,\phi)$ that can be obtained  by    considering the intersection of the horizontal lines   with the quadratic  black lines. We use the same colour code as in Fig.\ref{figura1}  for indicating the different energy regions. Dashed horizontal lines indicate benchmarking energies, ground-state energies, ESQPT energies  and $E/J=-1$ in the case of panel III. } 
\end{figure*}
%%%%%%%%%%%%%%%%%%%%%%%%%%%%%%%%%%%%%%

\section{Numerical methods}
\label{NumericalDet}

We wrote specialized codes in fortran90 to calculate the spectrum of the Hamiltonian (\ref{Ham}) and the Husimi representations. We performed the numerical diagonalization of the Hamiltonian (\ref{Ham}) by using the subroutine dsyev of the Lapack package \cite{lapack}. Once we had the eigenvalues and eigenstates we proceeded to calculate the Husimi representation (\ref{HusimiQ}) on a grid of canonical variables $Q$ and $P$. With the  eigenvalues and eigenstates of the Hamiltonian, we also computed the evolution of the initial coherent state (\ref{evolQ}), the survival probability (\ref{SPformula}) and  $\langle J\rangle(t)$ (\ref{eq:jzevol}).
For the entropy (\ref{EntropyW}) 
we used Monte Carlo sampling of the Husimi representation (\ref{HusimiQ}).

For the line integral (\ref{lineintegral}) along the classical trajectory, we divided the trajectory in small pieces, we evaluated (\ref{evolQ}) in each segment and finally we added the contributions from all circular segment i.e. numerical integration in polar coordinates.
In order to calculate  the classical survival probability and $j_z(t)$,  we  solved numerically the equations of motion corresponding to the Hamiltonian formalism in canonical variables $Q$-$P$ by using the software Mathematica.

\section{Energy density of states}
\label{AppA}
A classical approximation to the energy density of states (EDoS) can be obtained from the lowest order term of the Gutzwiller trace formula
\begin{equation}
\rho_{sc}(E)=\frac{J}{2\pi} \int dz d\phi  \   \delta \left(H(z,\phi)-E\right),
\end{equation}
where $H(z,\phi)$ is given in Eq. \ref{HclassicB}. To evaluate the previous integral, we use the properties of the Dirac delta to obtain
\begin{eqnarray}
\rho_{sc}(E)&=&\frac{1}{2\pi}\int_{\phi\in\mathbf{\Phi_\epsilon}}\frac{d\phi}{\sqrt{1-A(\phi)(2\epsilon-A(\phi))}}\times\nonumber\\
& &
\int_{-1}^{1} dz [\delta(z-z_+)+\delta(z-z_-)]\label{eq:appDen},
\end{eqnarray}
where
\begin{equation}
A(\phi)=\gamma_x\cos^2\phi+\gamma_y \sin^2\phi,
\label{eq:A}
\end{equation}
\begin{equation}
z_\pm(\phi,E)=\frac{1\pm\sqrt{1-A(\phi)[2\epsilon-A(\phi)]}}{A(\phi)},
\label{eq:Z}
\end{equation}
and $\mathbf{\Phi_\epsilon}$ is the set of values of $\phi$ for which there exist at least a solution of equation $H(z,\phi)/J=\epsilon$ for variable $z\in[-1,1]$, 
with $\epsilon=E/J$. In Fig. \ref{fig:App1}, we show $H(z,\phi)/J$ as a function of $z$ for different angles  $\phi$ and for the same four set of  couplings as  in Fig. \ref{figura1}. The plots allow not only to visualize the  range of allowed energies for given couplings, but also identify the different kind of trajectories that may appear in the  different  energy intervals. These latter  are indicated by the same color code used in Fig. \ref{figura1}. Observe that the number of intersections  of an  horizontal line (constant $\epsilon=E/J$)  with the quadratic curves $H(z,\phi)/J$ is equal to  the second integral in Eq.~(\ref{eq:appDen}): for the blue region in II, blue and orange regions  in III, and blue and green regions in IV, the number of intersection is two, while in the rest of regions this number is one.   
Equivalent expression for the EDoS were obtained in \cite{Ribeiro08} by analyzing the zeros of the eigenstates Husimi functions.

\section{Condition for real and avoided crossings}

 For sector III in parameter space and intermediate energy, there exist pairs of degenerate classical trajectories which are not connected by the parity symmetry of the model.      
One can obtain the set of  energy levels associated to these two disconnected  sets of classical orbits  by considering the Einstein-Birllouin-Keller(EBK)  quantization rule 
\begin{equation}
\frac{J}{2\pi}\int_{-\pi}^{\pi} \tilde{z}_\pm(\phi,E_{n_\pm}) d\phi= \left( n_{\pm}+\frac{1}{2}\right),    \label{EBKQ}
\end{equation}
where $n_\pm$ is an integer and  $\tilde{z}_\pm(\phi,E)$ are the two different classical trajectories for the same energy, $E$, expressed in terms of the canonical variables defined in Eq.~(\ref{eq:compon}).  These trajectories    are obtained from   the quantum Hamiltonian (\ref{HLMG}) by  mapping  the pseudospin operators to classical vectors $\hat{J}_i\rightarrow j_i$ and expressing $z$ as a function of $\phi$ and $\hat{H}\rightarrow E$, 
$$
\tilde{z}_\pm(\phi,E)=\frac{1}{\tilde{A}(\phi)}\pm \frac{\sqrt{1-\tilde{A}(\phi)[2\epsilon-\tilde{A}(\phi)]}}{\tilde{A}(\phi)},
$$
with $$\tilde{A}(\phi)=\frac{2J}{2J-1}\left(\gamma_x\cos^2\phi+\gamma_y \sin^2\phi\right)=\frac{2J}{2J-1}A(\phi)$$ and $\epsilon=E/J$. Notice that  $\tilde{z}$ and $             \tilde A$ are very similar to the functions defined in Eqs.~(\ref{eq:A}) and (\ref{eq:Z}).

If two energy levels   coincide $E_{n_+}=E_{n_-}$, by   summing  equations (\ref{EBKQ}) for both trajectories $\tilde{z}_\pm$, we obtain
$$
\frac{J}{\pi}\int_{-\pi}^{\pi} \frac{d\phi}{\tilde{A}(\phi)}= n_++n_-+1,
$$
which after performing the integral leads to $$
\frac{2J-1}{\sqrt{\gamma_x\gamma_y}}=n_++n_-+1.
$$
 This expression can be written  as
 $$
 \gamma_x\gamma_y= \left(\frac{2J-1}{p}\right)^2,
 $$
 where $p=n_++n_-+1>0$ is a positive  integer number.
 Since the intermediate regime with pairs of disconnected trajectories for the same energy appear only in the region $|\gamma_x|\geq 1$ and $|\gamma_y|\geq 1$,  the integer $p$ must be less or equal than $2J-1$. Then the condition for having crossing of  EBK levels can be written as 
 $$
 \gamma_x\gamma_y= \left(\frac{2J-1}{2J-N}\right)^2,
 $$
with $N$ an integer satisfying $0<N<2J$. We have verified that for $N$ odd  crossings between states of different parity appear, whereas for $N$ even  the crossing of the EBK energy levels become avoided crossings between states of same parity.

\section{Classical approximation to SP and $j_z$}
\label{AppB}
%%%%%%%%%%%%%%%%%%%%%%%%%%%%%%%%%%%%%%%
\begin{figure*}
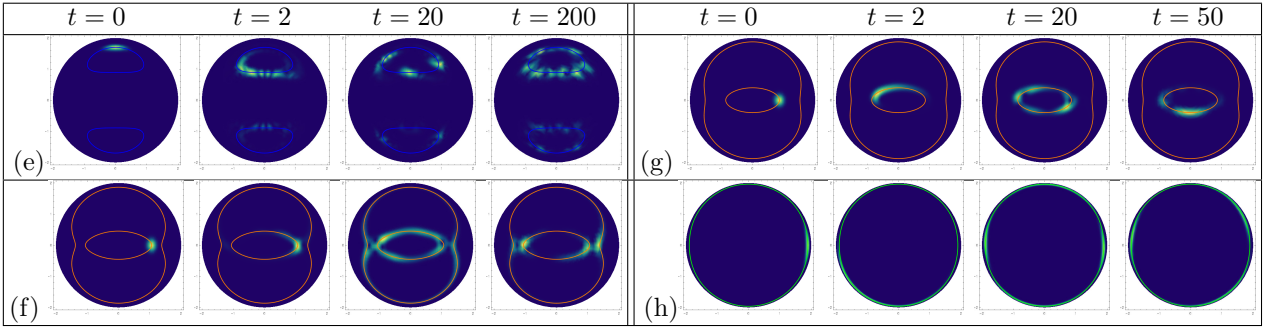

\begin{tabular}{|cccc||cccc|}
 %&
 \hline
 $t=0$ &  $t=2$ & $t=20$ & $t=200$ & $t=0$ &  $t=2$ & $t=20$ & $t=50$  \\ 
 \hline
(e)\includegraphics[width=1.8cm, height=1.8cm,angle=0]{figures/fig10e1.eps} &
 \includegraphics[width=1.8cm, height=1.8cm,angle=0]{figures/fig10e2.eps}  &
 \includegraphics[width=1.8cm, height=1.8cm,angle=0]{figures/fig10e3.eps} & \includegraphics[width=1.8cm, height=1.8cm,angle=0]{figures/fig10e4.eps}& 
 (g) \includegraphics[width=1.8cm, height=1.8cm,angle=0]{figures/fig10g1.eps} &
 \includegraphics[width=1.8cm, height=1.8cm,angle=0]{figures/fig10g2.eps}  &
 \includegraphics[width=1.8cm, height=1.8cm,angle=0]{figures/fig10g3.eps} &\includegraphics[width=1.8cm, height=1.8cm,angle=0]{figures/fig10g4.eps}\\
 \hline
(f) \includegraphics[width=1.8cm, height=1.8cm,angle=0]{figures/fig10f1.eps} &
 \includegraphics[width=1.8cm, height=1.8cm,angle=0]{figures/fig10f2.eps}  &
 \includegraphics[width=1.8cm, height=1.8cm,angle=0]{figures/fig10f3.eps} &\includegraphics[width=1.8cm, height=1.8cm,angle=0]{figures/fig10f4.eps}& (h)\includegraphics[width=1.8cm, height=1.8cm,angle=0]{figures/fig10h1.eps} &
 \includegraphics[width=1.8cm, height=1.8cm,angle=0]{figures/fig10h2.eps}  &
 \includegraphics[width=1.8cm, height=1.8cm,angle=0]{figures/fig10h3.eps} &\includegraphics[width=1.8cm, height=1.8cm,angle=0]{figures/fig10h4.eps}\\
 \hline
\end{tabular}
\caption{ \label{Husimitadditional} 
Density plots of the evolution of the Husimi functions $\mathcal{Q}_{\alpha_0}(\alpha,t)$ at selected times $t$ for four different initial Bloch coherent states. Solid orange line are  classical trajectory at the same energy of the respective initial state $\langle \alpha_o| \hat{H}|\alpha_o\rangle=H$. In all  panels  the coupling parameters are chosen at the avoided crossing condition $\gamma_x^{AC}=-4.10331$  (with $\gamma_y^{AC}=3\gamma_x^{AC}$ and $J=100$).  In panel (e) the initial state has an energy  well below the ESQPT. In (f) the initial state has an energy just above the ESQPT and is located in the inner trajectory of the degenerate pair. In  (g) the initial state has an energy in the intermediate regime but far enough from the ESQPT and is also located in  the inner trajectory of the degenerate pair. In (h) the initial state has an energy in the high energy regime well above $E/J=-1$.}
\end{figure*}
%%%%%%%%%%%%%%%%%%%%%%%%%%%%%%%%%%%%%%%%%%%

%%%%%%%%%%%%%%%%%%%%%%%%%%%%%%%%%%%%%%%%%%
\begin{figure*}
    \begin{tabular}{ccc}
     Profile & $ SP$ vs time & $ j_z $ vs time\\
   (e)     \includegraphics[width=5.2cm, height=3.2cm,angle=0]{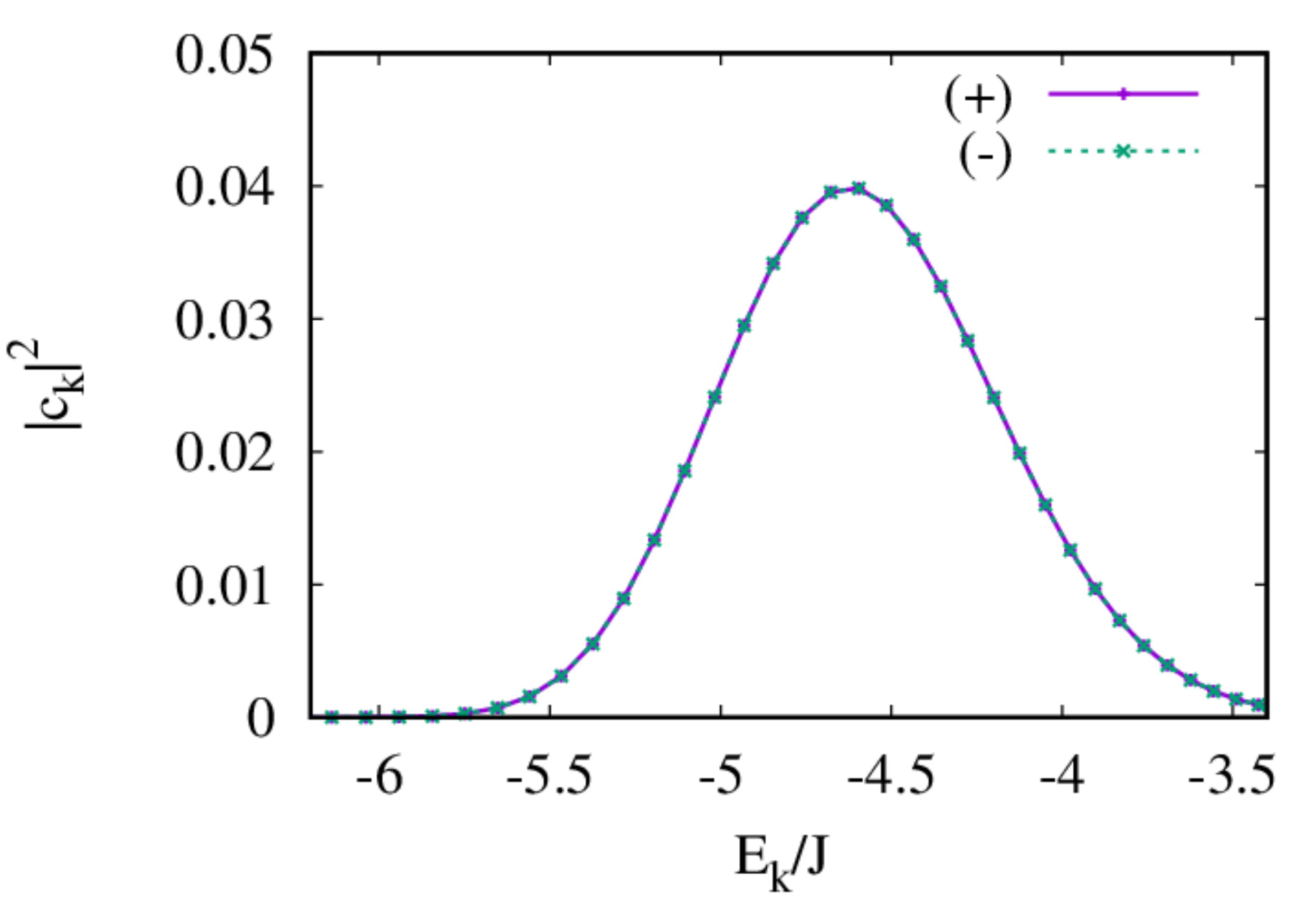} 
       &   \includegraphics[width=5.0cm, height=3.0cm, angle=0]{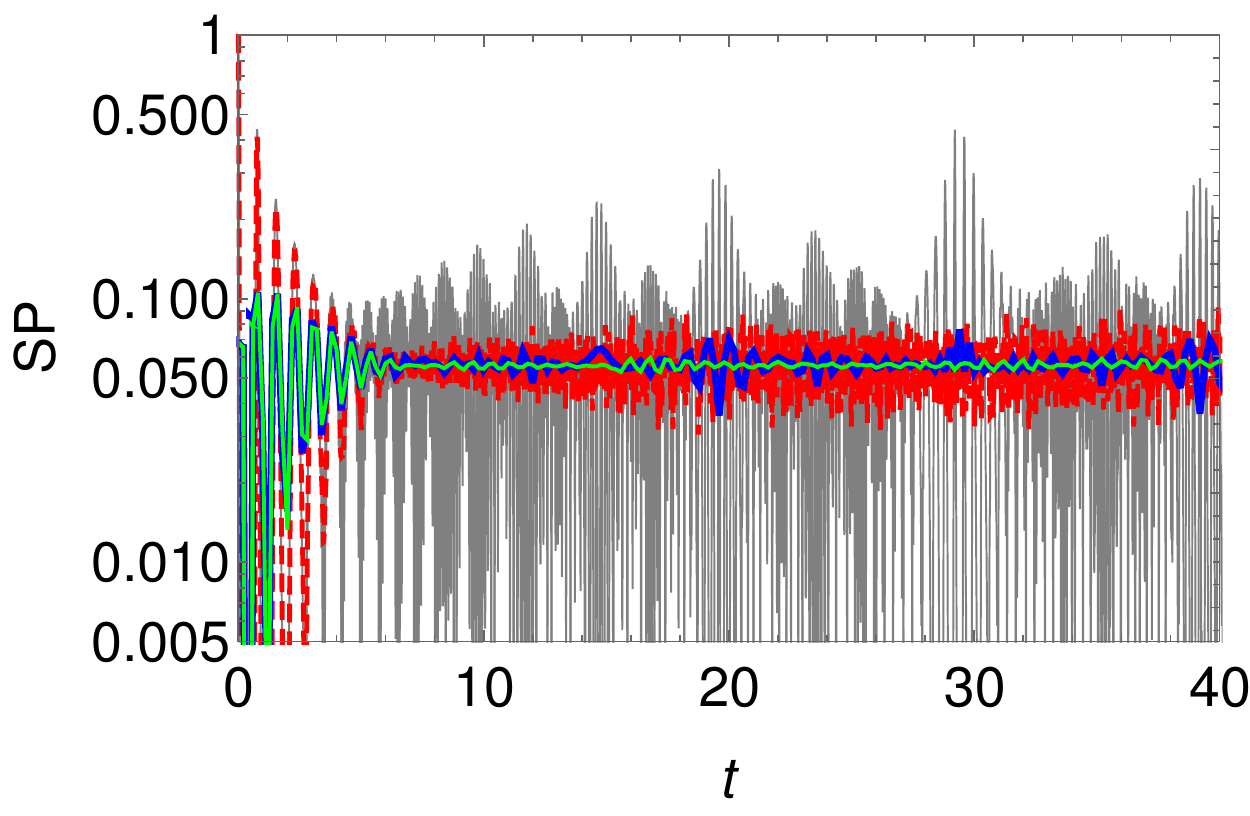}&\includegraphics[width=5.0cm, height=3.0cm, angle=0]{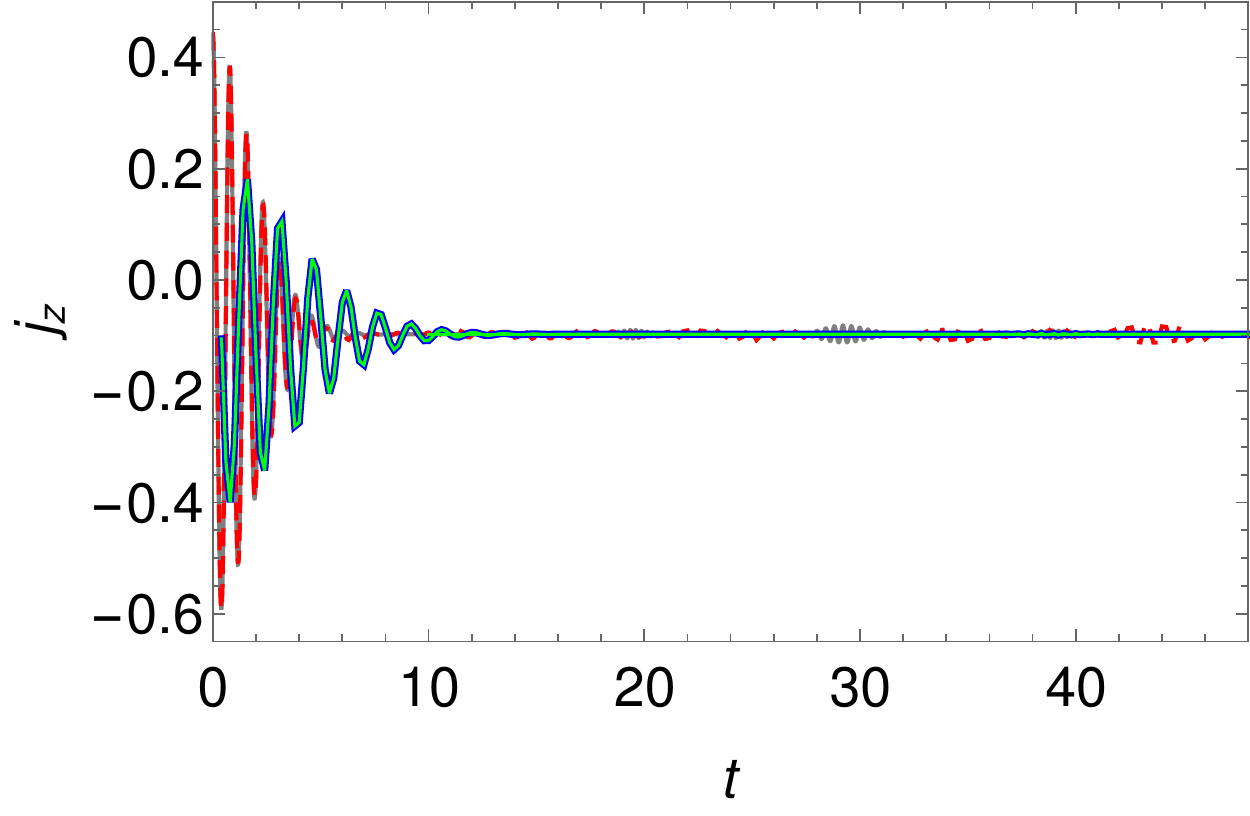}\\
       \hline
     (f) \includegraphics[width=5.2cm, height=3.2cm,angle=0]{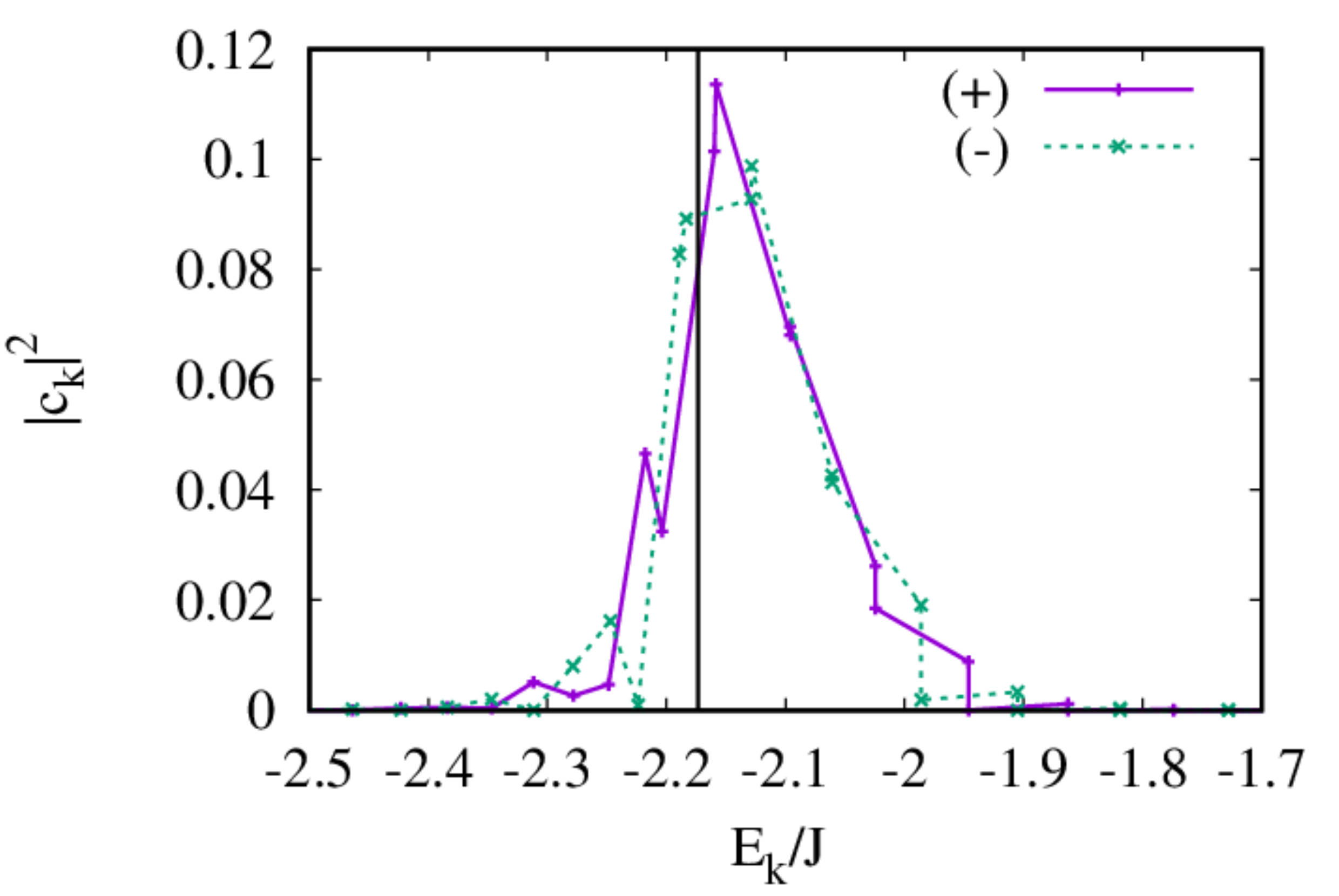} 
&\includegraphics[width=5.0cm, height=3.0cm]{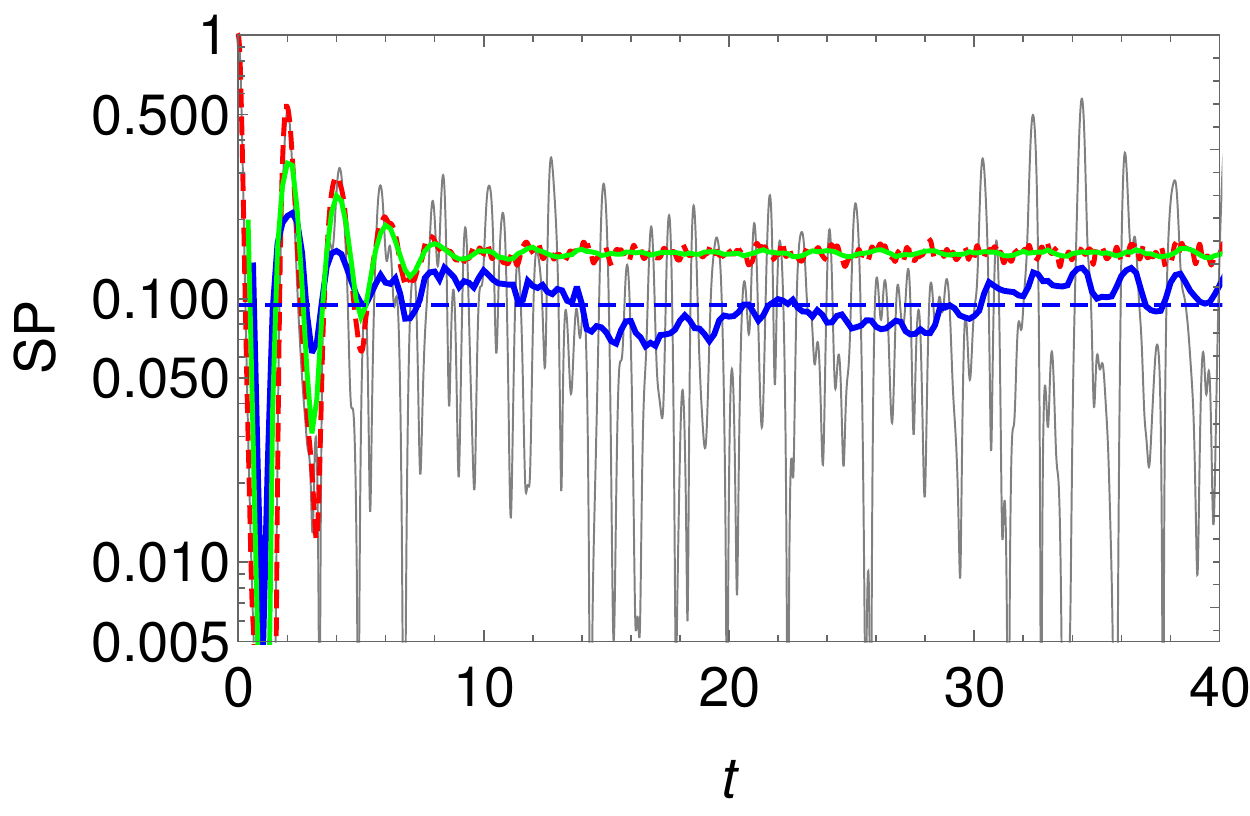}&\includegraphics[width=5.0cm, height=3.0cm, angle=0]{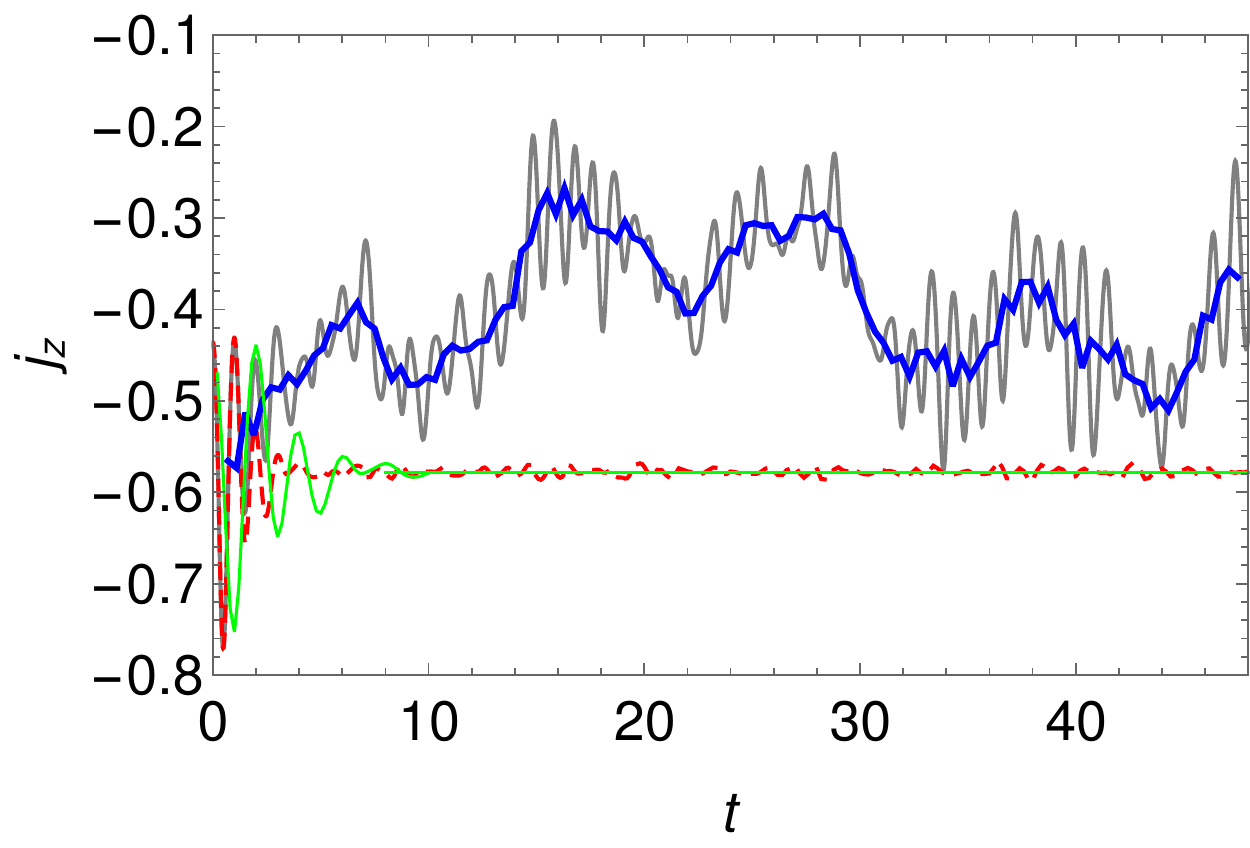} \\
\hline
(g) \includegraphics[width=5.2cm, height=3.2cm,angle=0]{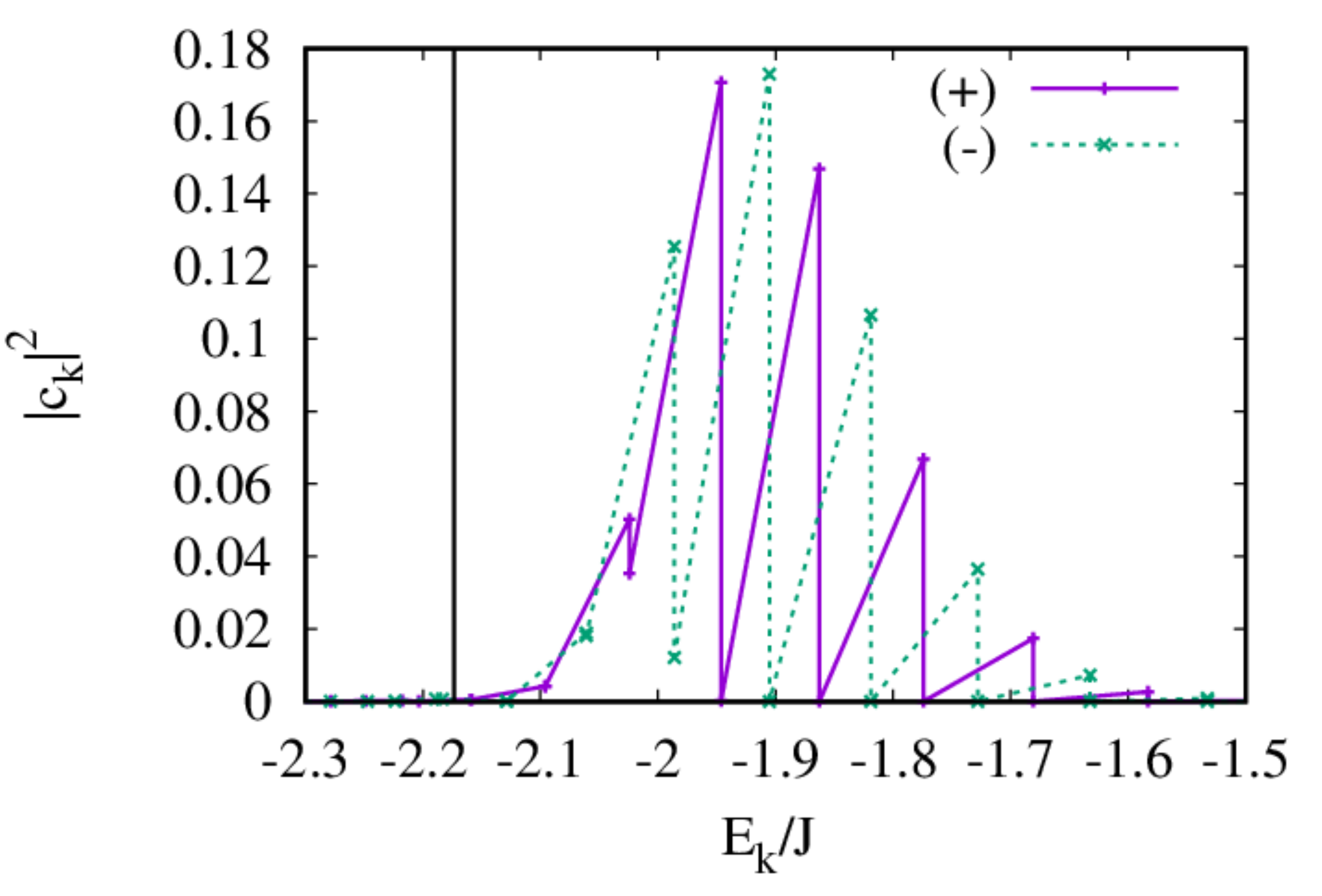} &
\includegraphics[width=5.0cm, height=3.0cm]{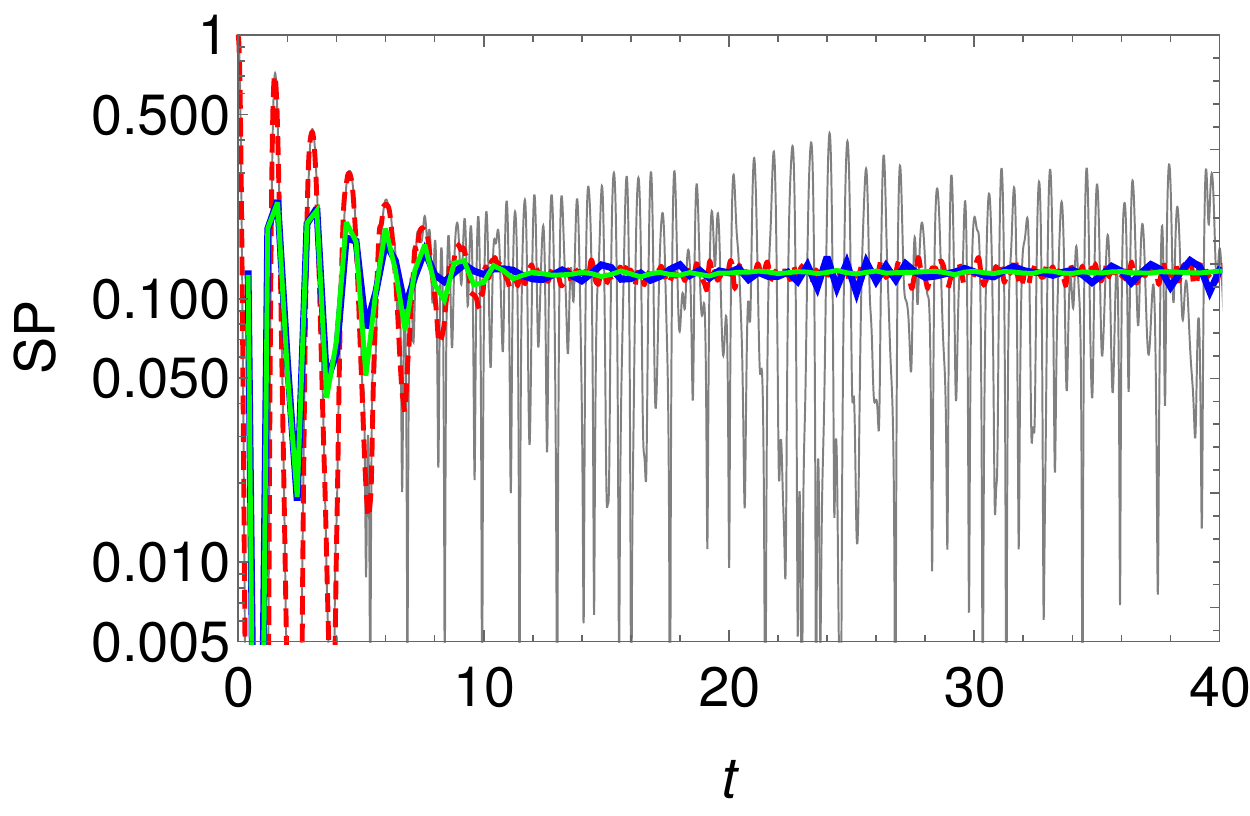}&
\includegraphics[width=5.0cm, height=3.0cm, angle=0]{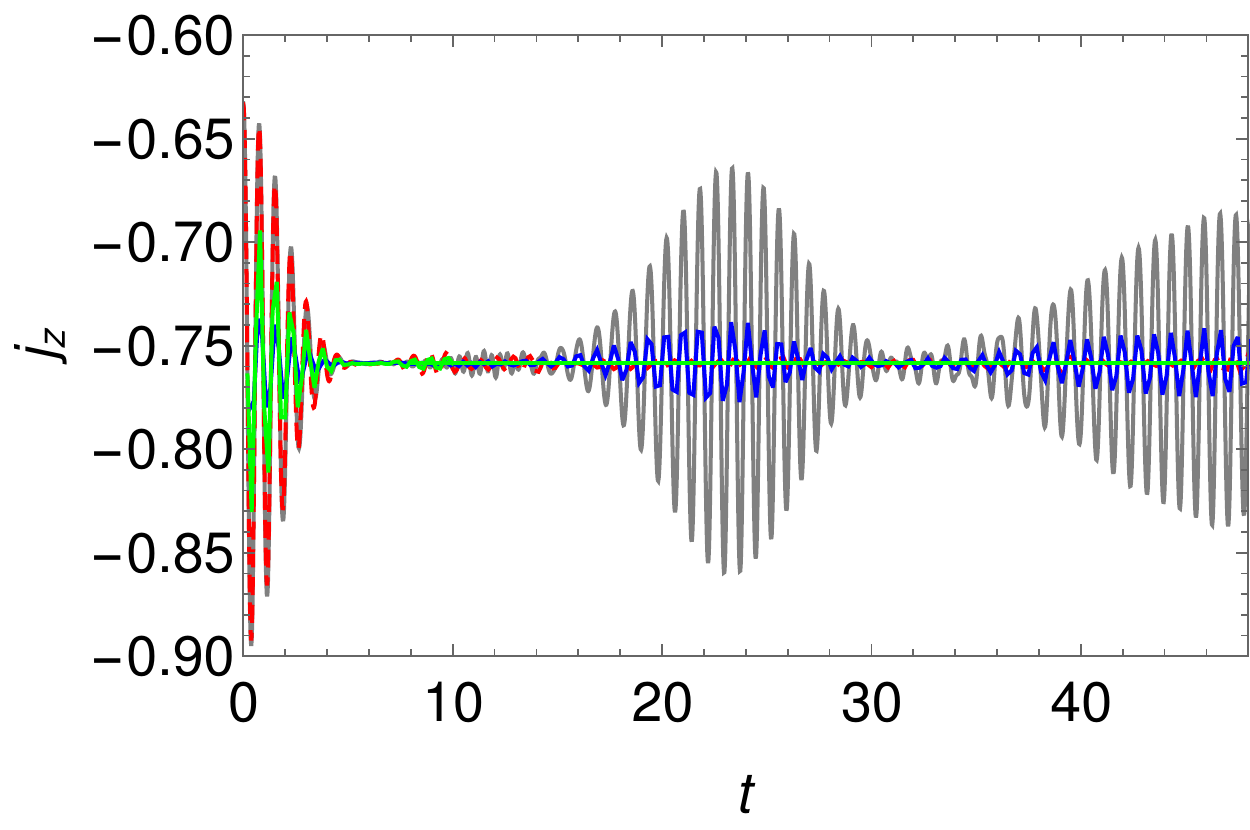} 
\\
\hline
(h)\includegraphics[width=5.2cm, height=3.2cm,angle=0]{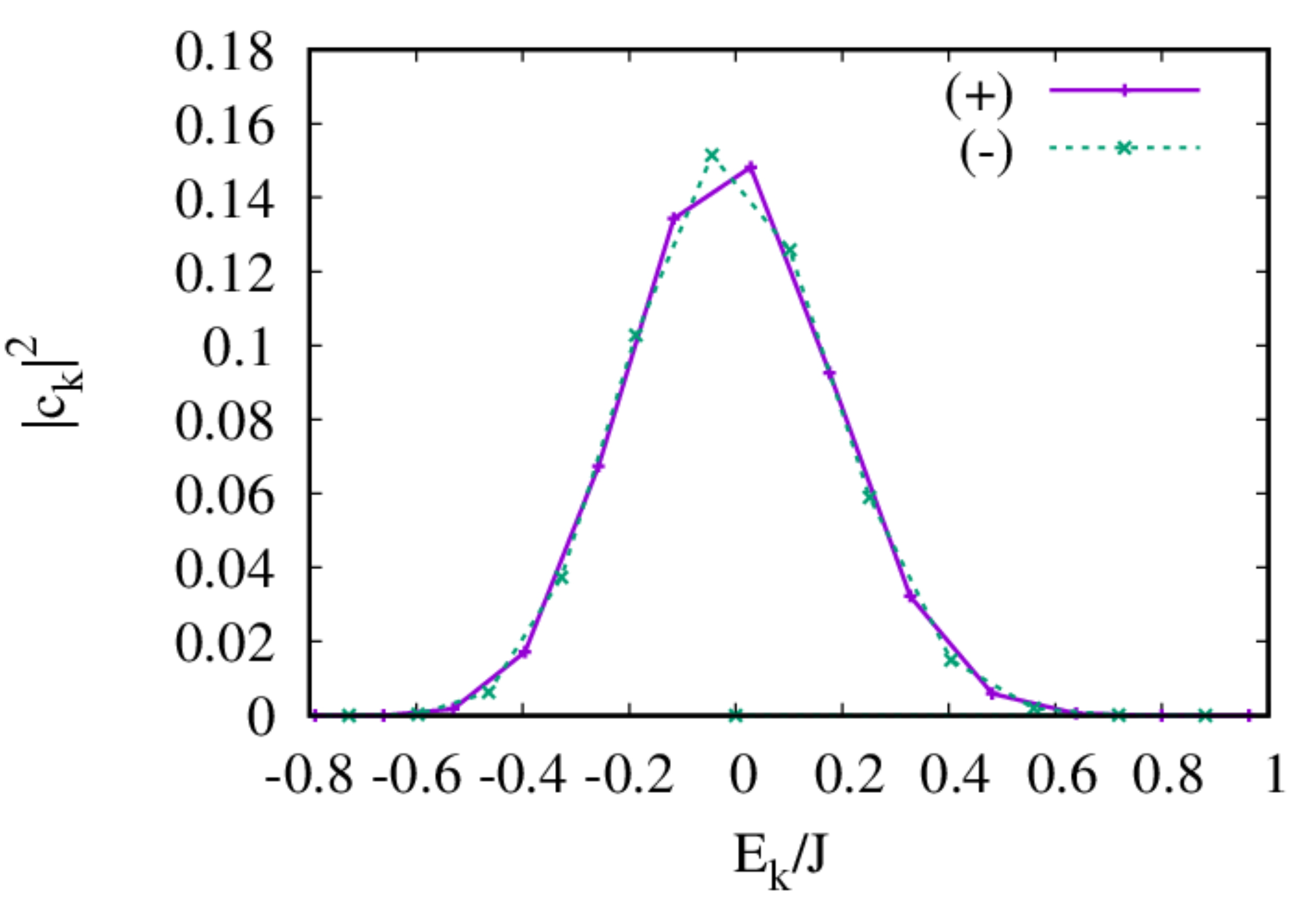}
&\includegraphics[width=5.0cm, height=3.0cm]{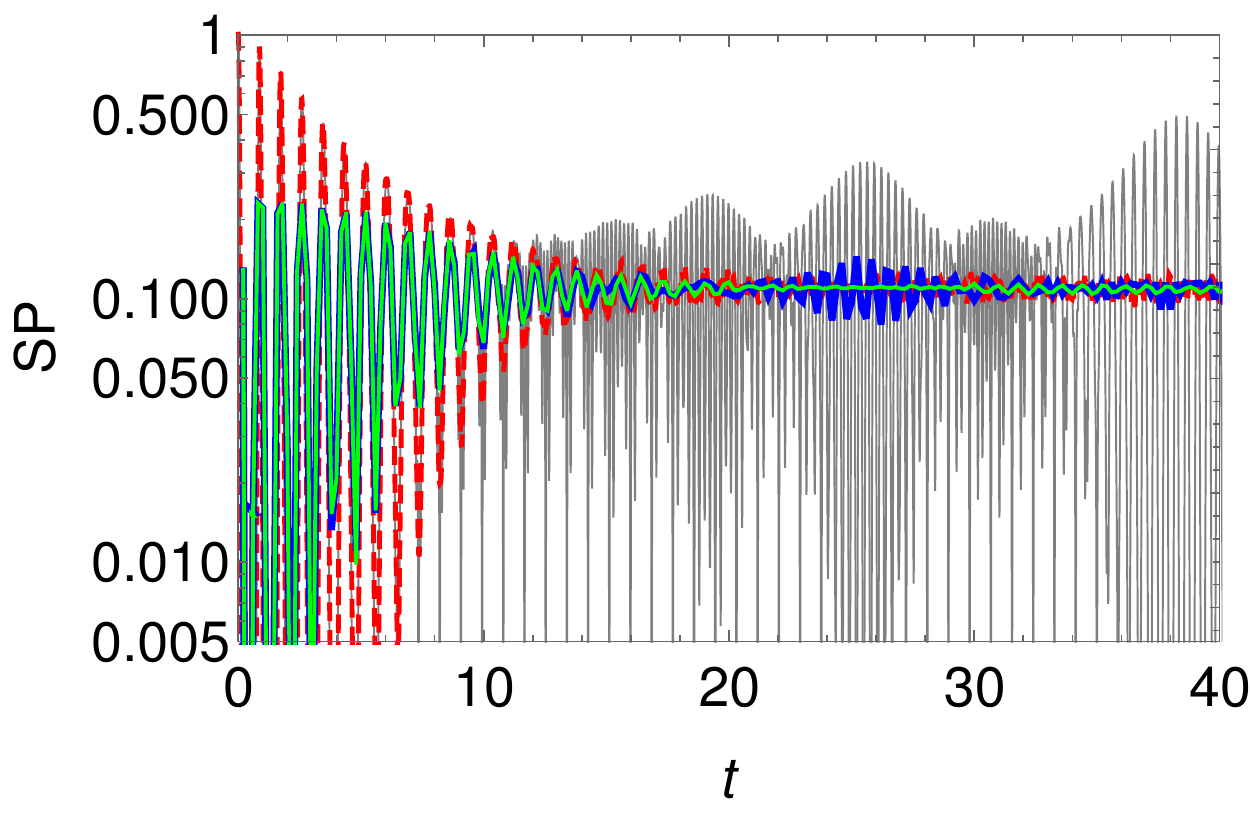}&\includegraphics[width=5.0cm, height=3.0cm, angle=0]{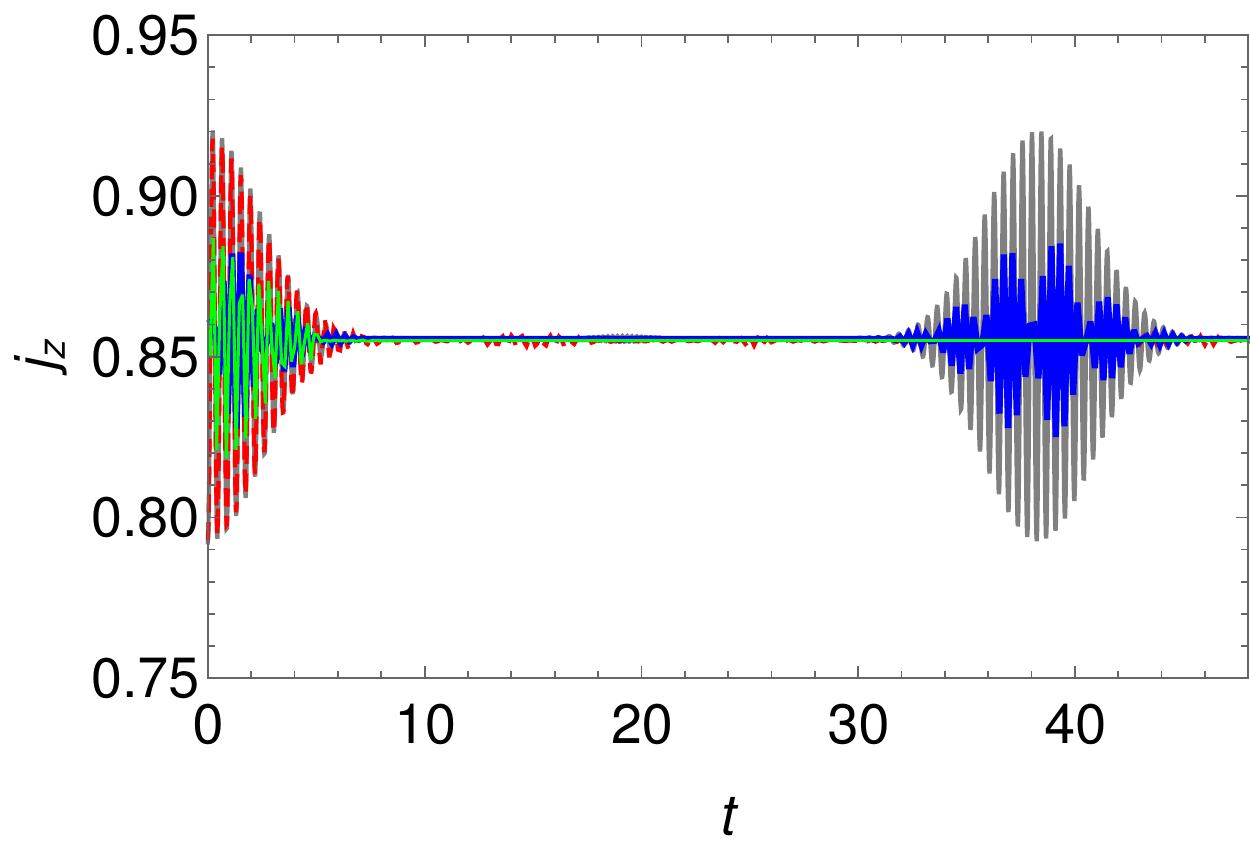} \\
\hline
    \end{tabular}
    \caption{\label{SPadditional}
    Same as Fig.~\ref{fig:8}, but for the four initial states of Fig.\ref{Husimitadditional}. Only in state (f) a breaking of the classical-quantum correspondence is observed in the rolling averages of observables $SP$ and $j_z$.  
    }
\end{figure*}
%%%%%%%%%%%%%%%%%%%%%%%%%%%%%%%%%%%%%%%%%%

\subsection{Survival probability}
Following Ref.\cite{Villasenor},  we express  the  survival probability of an initial Bloch coherent state $SP(t)=|\langle \alpha_o| \hat{U}(t)|\alpha_o\rangle|^2$  in terms of the Wigner functions at times $0$ and $t$,  
\begin{equation}
\label{SPwigner}
SP(t)=\frac{2\pi}{J} \int d{\bf u}\  w_{\alpha_o}({\bf u},0)w_{\alpha_o}({\bf u},t)\,,
\end{equation} 
where ${\bf u}=(Q,P)$ is a point in the %$J-$ scaled 
phase space. 
The  Wigner distribution for a Bloch coherent state is in turn given in terms of Legendre polynomials \cite{Yang_2019}
\begin{align}
\label{wignerformal}
&w_{\alpha_o}(\theta,\phi)=\\
&\frac{(2J)!}{4\pi}\sum_{k=0}^{2J}\sqrt{\frac{2k+1}{(2j-k)!(2j+k+1)!}}P_k(\cos\Theta)\,,\nonumber
\end{align}
where $\Theta$ is the angle between $\{\theta,\phi\}$ and $\{\theta_0,\phi_0\}$ i.e.
$$\Theta=\cos\theta\cos\theta_0+\sin\theta\sin\theta_0\cos(\phi-\phi_0)\,, $$
with  $\theta$ and $\phi$   angular spherical coordinates on the Bloch sphere.
For large $J$, the  Wigner distribution (\ref{wignerformal}) is well approximated by a normal distribution 
\begin{equation}
 \label{wigner}
 w_{\alpha_0}(\theta,\phi)\approx \frac{J}{\pi}e^{-J\Theta^2}\,.
\end{equation}
The Wigner function of the evolved state is obtained by  using  the truncated Wigner approximation (TWA),  which consists of assuming that the  evolution of the Wigner distribution is  governed by the classical equations of motion $\frac{\partial w}{\partial t}=\{w,H\}$. From this we obtain  
$
w({\bf u},t)\approx w(\varphi^{-t}({\bf u}),0)=w_{\alpha_o}(\varphi^{-t}({\bf u}))$, 
where $\varphi$ is the function which describes the classical time evolution 
${\bf u}(t)=\varphi^t({\bf u}_0)$ of an initial condition ${\bf u}_0$.
By using this result in the survival probability (\ref{SPwigner}) one gets the classical approximation,
\begin{eqnarray}
SP(t)&\approx&\frac{2\pi}{J} \int d{\bf u}\  w({\bf u},0)w(\varphi^{-t}({\bf u}),0)\nonumber \\
&=&\frac{2\pi}{J} \int d{\bf u}\  w_{\alpha_o}({\bf u})w_{\alpha_o}(\varphi^{-t}({\bf u})).\label{SPexp} 
\end{eqnarray}
The previous integral can be viewed as the average of the Wigner function at time $t$,  $w_{\alpha_o}^{(t)}$, weighted by the initial Wigner function $w_{\alpha_o}$,  or, equivalently, as the average of the initial Wigner function weighted by the Wigner function at time $t$ :  
\begin{equation}
\langle w_{\alpha_o}(\varphi^{-t}({\bf u}))\rangle_{w_{\alpha_o}}
=\langle w_{\alpha_o}({\bf u})\rangle_{w_{\alpha_o}^{(t)}}
\,.
\end{equation}
From this, the integral (\ref{SPexp}) can be estimated by means of Monte Carlo integration \cite{Villasenor}
\begin{equation}
 SP(t)\approx\frac{1}{M}\sum_{i=1}^M w_{\alpha_o}(\varphi^{\pm t}({\bf u}_i)),
\end{equation}
where ${\bf u}_i$  are
points randomly chosen following the initial distribution  $w_{\alpha_o}$ and $M$ is the total number of Monte Carlo iterations. %Equivalently the integral  can be estimated as
\subsection{Temporal evolution of $\langle \hat{J}_z\rangle$ }

The temporal evolution of the expectation value of the population operator can be calculated from the Wigner distribution
$$j_z(t)
\equiv\frac{1}{J}\langle \alpha_0|\hat{U}^\dagger(t)\hat{J}_z\hat{U}(t)|\alpha_0\rangle\,=
 \int d{\bf u}\  z({\bf u}) w(u,t),  
$$
where $z({\bf u})=-\cos\theta=\left(\frac{1}{2}(Q^2+P^2)-1\right)$.
Similar to the $SP$, we approximate classically this expression by using the TWA and the Gaussian distribution for the Wigner function of a coherent state~(\ref{wigner}),  
$$
j_z(t) \approx \int d{\bf u}
\ w_{\alpha_o}(\varphi^{-t}({\bf u})) z(\bf{u})\,.
$$
The previous integral can be interpreted as the average of $z(\bf{u})$ weighted by the Wigner function at time  $t$ ($w_{\alpha_o}^{(t)}$), 
$$
j_z(t)\approx \langle z({\bf u})\rangle_{w_{\alpha_o}^{(t)}}.
$$
To evaluate the integral, we also perform  a 
Monte Carlo integration to obtain
$$j_z(t) \approx\frac{1}{M}\sum_{i=1}^M z(\varphi^{ t}({\bf u}_i))\,,$$
where ${\bf u}_i$ is a set of $M$ random points generated from  the initial Wigner distribution (\ref{wigner}) and  which evolve according to   
the classical  Hamiltonian (\ref{Hclassic}).

\section{Additional cases}
\label{additional}

In this appendix we show the evolution of four different initial  Bloch coherent states, for   
couplings satisfying  the same  condition of avoided crossing as in the vertical solid line in Fig.\ref{figura2}(b).  In Fig.~\ref{Husimitadditional} the evolution of the Husimi function $\mathcal{Q}_{\alpha_0}(\alpha,t)$ is 
presented. States (e) and (h) are chosen with energies outside the intermediate energy regime where avoided crossings occur. State (e) has an energy well below the ESQPT, whereas state (h) has an energy well above $E/J=-1$. Initial  state  (f)  has an energy just above the ESQPT but, differently to state (a) in Fig.~\ref{lineintegrals},  it is initially located  on the inner trajectory and  tunneling takes place into the outer trajectory.  The initial state (g)  has an energy in the intermediate regime but far enough from the ESQPT,  its Husimi function is located  on the inner trajectory and no tunnelling is observed in its unitary evolution.   
 The squared energy  components of the initial states, their survival probability and the 
evolution of the expectation value $\langle \hat{J}_z \rangle$ are presented in Fig. \ref{SPadditional}. It is confirmed that only in the case of state (f) the classical-quantum correspondence is broken for the rolling averages of the considered observables.

\bibliography{biblio}
\end{document}